\documentclass[12pt]{article}
\usepackage{latexsym}
\usepackage{amsmath}
\usepackage{amssymb}
\usepackage{epsfig,graphics}
\usepackage{booktabs}
\usepackage{multirow}

\newcommand{\be}{\begin{equation}}
\newcommand{\ee}{\end{equation}}

\begin{document}

\begin{titlepage}

\vspace*{0.7in}

\begin{center}
{\large\bf Closed flux tubes in higher representations and their
  string \\ description in D=2+1 SU($N$) gauge theories\\}
\vspace*{0.7in}
{Andreas Athenodorou$^{a}$ and Michael Teper$^{b}$\\
\vspace*{.2in}
$^{a}$Department of Physics, Swansea University, Swansea SA2 8PP, UK\\
\vspace*{.1in}
$^{b}$Rudolf Peierls Centre for Theoretical Physics, University of Oxford,\\
1 Keble Road, Oxford OX1 3NP, UK
}
\end{center}

\vspace*{1.0in}

\begin{center}
{\bf Abstract}
\end{center}

We calculate, numerically,  the low-lying spectrum of closed confining flux tubes
that carry flux in different representations of SU($N$).
We do so for SU(6) at $\beta=171$, where the calculated low-energy physics is
very close to the continuum limit and, in many respects, also close to 
$N=\infty$. We focus on the adjoint,
$\underline{84}$, $\underline{120}$, $k=2A,2S$ and $k=3A,3M,3S$
representations and provide evidence that the corresponding flux
tubes, albeit mostly unstable, do in fact exist. We observe that 
the ground state of a flux tube with momentum along its axis
appears to be well defined in all cases and is well described by the
Nambu-Goto free string spectrum, all the way down to very small
lengths, just as it is for flux tubes carrying fundamental
flux. Excited states, however, typically show
very much larger deviations from Nambu-Goto than the corresponding
excitations of fundamental flux tubes and, indeed, cannot be
extracted in many cases. We discuss whether what we are seeing
here are separate stringy and massive modes or simply large 
corrections to energy levels that will become string-like at larger lengths.

\vspace*{0.95in}

\leftline{{\it E-mail:} a.a.athinodorou@swansea.ac.uk,  m.teper1@physics.ox.ac.uk}

\end{titlepage}

\setcounter{page}{1}
\newpage
\pagestyle{plain}

\tableofcontents

\section{Introduction}
\label{section_intro}

In the confining phase of SU($N$) gauge theories in $3+1$ or $2+1$
dimensions, the flux between 
sources in the fundamental representation is carried by a flux tube 
that at large separations, $l$,  will look like a thin string.
The spectrum of such a string-like flux tube,
whether closed (around a spatial torus) or open (ending at two sources),
should be calculable from an effective string action
\cite{LSW,PS} 
once  $l$ is large enough that the energy gap to the ground state 
has become small compared to the gauge theory's dynamical scale, 
$\sim O(\Lambda_{\overline{MS}})$ in $D=3+1$ and $\sim O(g^2)$ in $D=2+1$.
Indeed, it may be that the spectrum is simple even at smaller $l$, where
the energy gaps are large,  once $N$ is so large 
that flux tubes effectively do not mix or decay. In recent years
a great deal of progress has been made in determining the universal terms of
this effective string action thus determining the spectrum at large $l$. (See
\cite{OAZK}
for a recent review.)
Simultaneously, numerical lattice calculations have determined the spectrum at 
small to medium values of $l$, where the dynamics turns out to be remarkably 
close to that of a free string theory (Nambu-Goto in flat space-time). 

In this paper we extend our recent lattice calculations of the spectrum of
closed flux tubes in 2+1 dimensions
\cite{AABBMT_fd3}
to the case where the flux is in representations other than the 
fundamental. This will include cases where the flux tube is stable for all 
$l$ (e.g. the ground states of ${\cal{N}}$-ality $k=2$ or $k=3$) and cases 
where it is not. Whether the latter do have a well-defined identity is an
interesting question which we shall also address, albeit only
empirically in this paper. (We are not aware of a quantitative
theoretical analysis of the binding and decay of such `composite' flux tubes, 
although the general framework for decays has been developed in
\cite{decays},
and it would be interesting to understand if the flux tubes considered
in this paper satisfy the conditions for those calculations to be accurate. 
See also
\cite{decays2}
for related work.)  
As in our earlier work
\cite{AABBMT_fd3}
nearly all our calculations are in SU(6), where the theory is close to
its $N=\infty$ limit for many low-energy quantities, but far enough 
away from  that limit for the $k=2$ and $k=3$ ground states to be well
below their decay thresholds. Our calculations are at a fixed value of the 
lattice spacing $a$ that is small enough for most lattice corrections to be 
negligible (within our statistical accuracy). 

In the next Section we provide a (very) brief sketch  of relevant analytic and
numerical results. We then describe the technical aspects of the lattice 
calculation. In Section~\ref{section_results} we present our
results. We begin with flux tubes carrying flux in the $k=2$ symmetric and
antisymmetric representations (that arise from $f\otimes f$, where $f$ is the
fundamental representation), then move on to the three minimal $k=3$
representations (arising from $f\otimes f\otimes f$), the adjoint flux tube
(from  $f\otimes {\bar{f}}$) and those carrying flux 
in the  \underline{84} and \underline{120} representations
(arising from $f\otimes f\otimes {\bar{f}}$). The Appendix describes the 
properties of these representations.  Such flux 
tubes, when they exist, can be thought of as bound states of 
(anti)fundamental flux tubes and their spectra should contain the imprint
of the  massive modes associated with that binding. The latter should be
additional to the usual massless stringy modes, which are the only ones
to  appear in the spectrum of fundamental flux tubes in $D=2+1$
\cite{AABBMT_fd3}.

The lattice calculations are very similar to our earlier work with fundamental
flux and we refer to that work
\cite{AABBMT_fd3}
for most of the technical details. We also note our earlier calculation
of the spectrum of $k=2$ flux tubes
\cite{AABBMT_k2d3}
performed at smaller $N$ and for coarser $a$, and to earlier calculations of
$k$-string tensions
\cite{BBMT_kd3}. 
We refer to these for a more detailed discussion of $k$-strings.

\section{Background and Overview}
\label{section_background}

We are interested in the spectrum of flux tubes that are closed around
a spatial torus of length $l$. We make the sizes of the transverse spatial torus,
$l_\perp$, and the (Euclidean) temporal torus, $l_t$, large enough
that the resulting finite size corrections are negligible. As $l$
decreases, the theory suffers a finite volume transition at
$l=l_c=1/T_c$ where $T_c$ is the deconfining temperature,
and for $l\leq l_c$ the theory does not support winding
flux tubes. This
transition is strongly first order for SU(6), the case of interest in
this paper. Since $T_c \sim \surd\sigma_f$ in terms of the fundamental
string tension, this means we can study closed flux tubes of length 
$l \gtrsim 1/\surd\sigma_f$. 

Since the spectrum of the Nambu-Goto model turns out to be an excellent
starting point for much of the observed fundamental flux tube spectrum, 
we begin by briefly summarising it. We then say something about relevant
analytic results for long flux tubes -- an area in which striking
progress has been made in the last few years -- as well as the
numerical results for flux tubes in the fundamental representation, 
which the present work extends to higher representations. We then 
say something about those higher representations.

\subsection{analytic expectations}
\label{subsection_analytic}

Recall that we consider flux tubes that are closed around a spatial torus of length $l$,
with the transverse and Euclidean time tori chosen so large as to be 
effectively infinite. Such flux tubes may carry non-zero longitudinal
momentum.  (We do not consider non-zero momentum transverse to 
the string since that does not teach us anything new.)
In the $N\to\infty$ limit where decays and mixings 
are suppressed, the world sheet swept out by the propagating flux tube
has no handles or branchings and so has the simple topology of a cylinder. 
The simplest effective string action is proportional to the invariant area 
of the sheet in flat space-time (Nambu-Goto). The Nambu-Goto spectrum 
arises from left and right moving massless `phonons' on the background 
string of tension $\sigma$. Let $n_{L(R)}(k)$ be the number of left(right) 
moving phonons of momentum $|p|=2\pi k/l$ and define their
total energy to be $2\pi N_{L(R)}/l$, i.e.
\begin{equation}
N_L = \sum_k  n_L(k)k, \qquad
N_R = \sum_k  n_R(k)k.
\label{eqn_NLR}
\end{equation}
so that the state has total longitudinal momentum $p=2\pi q/l$ with
\begin{equation}
N_L - N_R = q.
\label{eqn_NLRmom}
\end{equation}
The energy levels in $D=2+1$ turn out to be given by
\cite{Arvis,LW}
\begin{equation}
E_n(q,l)
=
\left\{
(\sigma l)^2 
+
8\pi\sigma \left(\frac{N_L+N_R}{2}-\frac{1}{24}\right)
+
\left(\frac{2\pi q}{l}\right)^2\right\}^{\frac{1}{2}}
\label{eqn_EnNG}
\end{equation}
where $n=N_L+N_R$ and one can readily calculate the degeneracy
of a given energy level. 
We note that the parity of a state is given by
\begin{equation}
P = (-1)^{number\ of\ phonons}.
\label{eqn_Pd3}
\end{equation}
We display in Table~\ref{table_NGstates} the states which we will 
later discuss in more detail. (Here $a_{\pm k}$ creates a phonon
of momentum $\pm 2\pi k/l$.)
\begin{table}[t]
\begin{center}
\centering{\scalebox{1.02}{\small
\begin{tabular}{|c||c|c||c|}  \hline \hline
\ \ \ \ \ \ \ \ \ \ \ \ \ \ \ \ $N_L,N_R$ \ \ \ \ \ \ \ \ \ \ \ \ \ \
\ \ & \ \ \ \ $q$ \ \ \ \ & \ \ \ \ $P$ \ \ \ \  & \ \ \ \ \ \ \ \ \ \
String State \ \ \ \ \ \  \ \ \ \ \\ \hline \hline
 $N_L=0,N_R=0$ & $0$ &  $+$ & $| 0 \rangle$ \\ \hline
 $N_L=1,N_R=0$ & $1$ &  $-$ & $a_{1}| 0 \rangle$ \\ \hline
$N_L=1,N_R=1$ & $0$ &  $+$ & $a_{1} a_{-1} | 0 \rangle$ \\ \hline
\multirow{2}*{$N_L=2,N_R=0$} & \multirow{2}*{$2$} &  $+$ & $a_{1} {a}_{1} | 0 \rangle$ \\
&  &  $-$ & $a_{2} | 0 \rangle$ \\ \hline
\multirow{2}*{$N_L=2,N_R=1$} & \multirow{2}*{$1$} &  $+$ & $a_{2} a_{-1}| 0 \rangle$ \\
&  &  $-$ & $a_{1} a_{1} a_{-1} | 0 \rangle$ \\\hline
\multirow{4}*{$N_L=2,N_R=2$} & \multirow{4}*{$0$} &  $+$ & $a_{2} a_{-2} | 0 \rangle$ \\
&  &  $+$ & $a_{1} a_{1} a_{-1} a_{-1} | 0 \rangle$ \\
&  &  $-$ & $a_{2}a_{-1} a_{-1} | 0 \rangle$ \\
&  &  $-$ & $a_{1}a_{1} a_{-2}  | 0 \rangle$ \\\hline
\multirow{9}*{$N_L=3,N_R=3$} & \multirow{9}*{$0$} &  $+$ & $a_{3} a_{-3} | 0 \rangle$ \\
&  &  $+$ & $a_{2} a_{1} a_{-2} a_{-1} | 0 \rangle$ \\
&  &  $+$ & $a_{1} a_{1} a_{1} a_{-1} a_{-1} a_{-1} | 0 \rangle$ \\
&  &  $+$ & $a_{1} a_{1} a_{1} a_{-3} | 0 \rangle$ \\
&  &  $+$ & $a_{3} a_{-1} a_{-1} a_{-1} | 0 \rangle$ \\
&  &  $-$ & $a_{3} a_{-2} a_{-1} | 0 \rangle$ \\
&  &  $-$ & $a_{2} a_{1} a_{-3} | 0 \rangle$ \\
&  &  $-$ & $a_{2} a_{1} a_{-1}  a_{-1}  a_{-1}| 0 \rangle$ \\
&  &  $-$ & $a_{1} a_{1} a_{1} a_{-2} a_{-1} | 0 \rangle$ \\\hline
\end{tabular}}}
\end{center}
\vspace{-0.5cm}
\caption{\label{table_NGstates}
The states of the lowest Nambu-Goto energy levels with $p=2\pi q/l$
for $q=0,1,2$, and $q=0$ excited states with $N_L+N_R \le 6$.}
\end{table}
We refer to
\cite{AABBMT_fd3}
for a more detailed discussion, and  reasons why we ignore quantum numbers
other than parity and momentum along the flux tube.

Note that the spectrum in eqn(\ref{eqn_EnNG}) is derived using naive 
light-cone quantisation
\cite{Arvis}. 
Its actual relationship with Nambu-Goto in
$D < 26$ is a subtle question, which is considered critically in 
\cite{OAZK}.
We will nonetheless use it for comparative purposes and refer to it as `the
Nambu-Goto spectrum'.

For large enough $l$ we can expand eqn(\ref{eqn_EnNG}) in powers of
$1/\sigma l^2$. The first correction to the linear $\sigma l$ piece 
coincides with the well-known $O(1/l)$ universal L\"uscher correction
\cite{LSW,PS}.
 It is now known that the $O(1/l^3)$ correction is also universal
\cite{LW,JM}
as is the  $O(1/l^5)$ correction
\cite{OA}. 
(This is for $D=2+1$; there are interesting differences in $D=3+1$
\cite{OAZK}). 
The universality class is determined by the massless modes living on the
string. If, as is plausible here, the only such modes are those
arising from the bosonic massless transverse oscillations, then these
universal terms coincide with the corresponding terms in the expansion
of the Nambu-Goto action and energy levels
\cite{OA,FG}. 
Thus once $l$ is large
enough for the expansion of eqn(\ref{eqn_EnNG}) in powers of $1/l^2\sigma$
to converge (which occurs at small $l$ only for the absolute ground state) 
we can expect the free string Nambu-Goto theory to provide an 
increasingly accurate description of that part of the closed flux tube
spectrum. 

Note that such effective string calculations become valid for an
excited flux tube once $l$ becomes large enough that the energy 
gap  $\Delta E_n \simeq E_n(l) - \sigma l$ becomes small compared to 
the dynamical energy scale of the theory $\sim \surd\sigma$. And this
is so independent of $N$. However the expansion of eqn(\ref{eqn_EnNG})
only requires $\Delta E_n \lesssim \sigma l$ which is a weaker condition.
So for the effective string approach to be valid all the way down to the 
Nambu-Goto radius of convergence we presumably need to invoke nearness 
to the $N=\infty$ limit as well. 

While the above analytic progress has so far concerned flux tubes at 
large enough $l$, we remark that there have been promising recent attempts at 
understanding the spectrum at smaller $l$ from considerations of the 
scattering matrix of phonons on the world sheet 
\cite{SDRFVG_13}.
(See also
\cite{SDRFVG_12}.) 

These analytic results assume that the flux is carried by a single
flux tube. While this is indeed the case for fundamental flux tubes,
in appropriate limits, it is not clear what happens for higher 
representations $R$. While we can still expect an effective (Goldstone)
action approach to be valid as long as  
$\Delta E_n \simeq E_n(l) - \sigma_R l <\surd\sigma_f$, extending
the range of validity by an appeal to large $N$ is dubious. Indeed, 
in the $N=\infty$ limit we expect the flux to be carried by an
appropriate number of non-interacting fundamental flux tubes.
(As we shall see below when we consider explicit operators for such
flux.) Thus we are not able to rely on an ideal $N\to\infty$ limit in
the same way as we can for fundamental flux tubes.

\subsection{fundamental flux tubes}
\label{subsection_fundamental}

In
\cite{AABBMT_fd3}
we performed calculations in SU(6) of the closed flux tube spectrum on the 
same lattices, and at the same coupling as in this paper. We briefly list
some of the conclusions of that work that are relevant to this paper. \\
1) The absolute ground state is very accurately described by the free string
prediction in eqn(\ref{eqn_EnNG}), with a correction only becoming visible for
$l\surd\sigma_f \lesssim 2$. \\
2) This correction is consistent, within the errors, with being either 
$\propto 1/l^5$ or  $\propto 1/l^7$, where the latter is the prediction 
of the analysis of universal terms. \\
3) The lightest states with $p\neq 0$ also show no visible correction to Nambu-Goto 
down to  $l\surd\sigma_f \sim 1.5$. These states contain phonon excitations
and so we see that the flux tube behaves like an excited thin string even when its 
length is about the same as its width (which is naively $\sim   \surd\sigma_f$).\\
4) In general whenever an excited state corresponds to phonons that are all right
or left moving, corrections to Nambu-Goto are almost invisible. \\
5) While other low-lying excited states typically show larger corrections, these 
typically become insignificant at values of $l$ that are much smaller 
than required for the expansion of eqn(\ref{eqn_EnNG}) in powers of $1/l^2\sigma$ 
to become convergent. 
That is to say, our results show that the Nambu-Goto prediction is 
still good when all the terms in the  $1/l^2\sigma$ expansion are
important: i.e. the series of correction terms must itself
resum to a modest total correction even at small $l$. \\
6) There is no evidence at all of any non-stringy massive modes that are additional 
to the stringy ones that are well described by the free string theory spectrum.

One of our main motivations for the present study is to contrast the above with what one
finds for flux tubes that are bound states of fundamental flux tubes, and where
the binding, measurable through the value of the string tension, provides unambiguous 
massive dynamics that should somehow make itself seen in the flux tube spectrum.

\subsection{flux tubes in higher representations}
\label{subsection_highrep}

Consider two well separated sources in representations $R$ and $\overline{R}$. The flux
between them will be carried by one or more confining flux tubes and,
if we ignore the possibility of screening, will be in the
representation $R$. 

The representations of SU(6) that 
we consider in this paper are the fundamental $f$, the adjoint $A$ 
which appears in $f \otimes \overline{f}$, the representations $\underline{84}$
and    $\underline{120}$ which appear in  $f \otimes f \otimes \overline{f}$,
and the various irreducible representations generated by $f \otimes f$ 
and $f \otimes f \otimes f$. These last two
belong to the $k=2$ and $k=3$ sectors respectively. That is to say under a
global gauge transformation that is an element of the centre, 
$g(x)=e^{i\pi/N}\mathbb{I}$, the sources transform as 
$=e^{i\pi k/N}$. Under this categorisation the $f$, $\underline{84}$ and  
$\underline{120}$ belong to $k=1$ and the adjoint $A$ to $k=0$. In the $k=2$ sector we
consider the antisymmetric $2A$ and symmetric $2S$ representations.
In the $k=3$ sector we consider the antisymmetric $3A$, the mixed $3M$, 
and the symmetric $3S$ representations. All these representations are
discussed in more detail in the Appendix. 

Since gluons transform trivially under the centre, screening cannot 
change the value of $k$. Hence the absolute ground state in each $k$-sector 
will correspond to an absolutely stable flux tube. These are often referred to
as $k$-strings, although this term is often used more loosely to label all 
states in a given $k$-sector. Note that there will be an absolutely stable ground 
state for each parity, $P$, and longitudinal momentum, $p$, within each 
$k$-sector. (Note also that at a given $l$ the lightest state with
such non-trivial quantum numbers may include a glueball that carries 
some of the quantum numbers. Such states decouple from our calculations
in the $N\to\infty$ limit and, as we shall see, appear to play no role even
for $N=6$.) 

Earlier work 
\cite{BLMT_strings,BLMTUW_imp,AABBMT_k2d3}
has shown that the $k=2,3$ ground states are almost exactly $2A$ and $3A$ 
respectively, except when the flux tube is very short, $l\sim l_c$. This is 
related to the observation that, despite the fact that gluon screening can take one
from e.g. $2S$ to $2A$, the actual overlap is found to be extremely small
\cite{BBMT_kd3}.
This interesting feature of the dynamics is something we shall examine in more 
detail in this paper. 

We note that some overlaps are lower order in $1/N$ and hence would be naturally
suppressed for SU(6). This includes the overlap of the adjoint flux tube to the 
vacuum (or glueballs) and  the $\underline{84}$ and $\underline{120}$ flux tubes
to a single fundamental, $f$, flux tube. 
On the other hand the overlap of the adjoint onto
a pair of flux tubes, one $f$ and the other $\overline{f}$, should not be 
suppressed. Similarly for $\underline{84}$ and $\underline{120}$ to three
flux tubes, 2 $f$s and one $\overline{f}$. We will be careful to discuss
these possibilities when we present our results below.

\section{Lattice methods}
\label{section_lattice}

\subsection{lattice setup}
\label{subsection_setup}

Our space-time is a periodic cubic 
$L_x\times L_y\times L_t$ lattice with lattice spacing $a$. The degrees 
of freedom are SU($N$) matrices, $U_\mu(x,y,t)$ or more compactly $U_l$, 
assigned to the links $l$ of 
the lattice. The action is the standard Wilson plaquette action,
so the partition function is 
\begin{equation}
Z(\beta)
=
\int \prod_l dU_l \,
e^{-\beta \sum_p\{ 1 - \frac{1}{N}{\mathrm{ReTr}}U_p\}}
\label{eqn_latticeZ} 
\end{equation} 
where $U_p$ is the ordered product of matrices around the
boundary of the elementary square (plaquette) labelled by $p$.
Taking the continuum limit, one finds that
\begin{equation}
\beta\stackrel{a\to 0}{=}\frac{2N}{a g^2}
\label{eqn_beta} 
\end{equation} 
where $g^2$ is the coupling and $a g^2$ is the dimensionless coupling
on the length scale $a$. The continuum limit is approached by 
tuning $\beta = 2N/ag^2 \to \infty$.

\subsection{calculating energies}
\label{subsection_energies}

 Here we give a brief sketch and refer the reader to Section 3 of
\cite{AABBMT_fd3}
for a detailed exposition.

We calculate energies from the time behaviour of correlators of
suitable operators $\{\phi_i\}$,
\begin{equation}
C_{ij}(t) 
=
\langle \phi_i^\dagger(t)\phi_j(0) \rangle
=
\langle \phi_i^\dagger e^{-Han_t} \phi_j \rangle
=
\sum_k c_{ik} c^\star_{jk} e^{-aE_k n_t}.
\label{eqn_cortomass}
\end{equation}
Since we wish to project onto loops of flux closed around the
$x$-torus, we use operators that wind around the $x$-torus.
The simplest such operator is the Polyakov loop
\begin{equation}
l_p(n_y,n_t) = \prod^{L_x}_{n_x=1} U_x(n_x,n_y,n_t)
\quad ; \quad
\phi(n_y,n_t) = \mathrm{Tr}_R \{l_p(n_y,n_t)\}
\label{eqn_poly}
\end{equation}
where $l=aL_x$ (we shall measure $l$ in physical units and $L$ in lattice
units, unless indicated otherwise) and we have taken the product of
the link matrices in 
the $x$-direction, around the $x$-torus and the trace is taken in the
desired representation $R$. We also use many other winding paths, as listed 
in Table 2 of
\cite{AABBMT_fd3},
and also with smeared and blocked SU($N$) link matrices
\cite{AABBMT_fd3}.
Using all these paths we can project onto different longitudinal momenta
and parities. The transverse momentum dependence is determined by Lorentz
invariance and so we only consider $p_\perp=0$ operators, obtained by summing 
over spatial sites, e.g. $l_p(p_\perp=0,n_t) \propto \sum_{n_y} l_p(n_y,n_t)$ 
in eqn(\ref{eqn_poly}). Unless otherwise stated all winding operators
in this paper will be with   $p_\perp=0$.

We now perform a variational calculation of the spectrum,
maximising $ \langle e^{-Ht} \rangle$  over this basis 
(usually projected onto the desired quantum numbers) . We usually do
so for $t=a$ and this provides us with an ordered set of approximate
energy eigenoperators  $\{\psi_i\}$. We then form the correlators
of these, $\langle \psi^\dagger_i(t) \psi_i(0) \rangle$, and 
extract the energies from plateaux in the effective energies, 
defined by
\begin{equation}
\frac{\langle \psi^\dagger_i(t) \psi_i(0) \rangle}
{\langle \psi^\dagger_i(t-a) \psi_i(0) \rangle}
=
\exp\{-aE_{i,eff}(t)\}.
\label{eqn_Eeffpsi}
\end{equation}
These plateaux typically
begin at values of $t$ that are larger than $t=a$. Given the
propagation of statistical errors, we can only identify such a plateau
if it corresponds to the operator having a large overlap onto the
desired state. Note that this largely excludes the possibility that our 
energy estimate is
contaminated by a small admixture of a lower lying state. (The effective
energy only provides an upper bound on the desired energy if extracted
where we perform the variational calculation, i.e. $t=a$ in our case.) It
is only where we have significant evidence for a plateau that we quote
an energy.

This above procedure is appropriate for stable states. However many of
our states will be unstable. (We will usually indicate that
in our figures.) If these states are  analogous to narrow resonances
then they are just as relevant to us as they would be if stable. 
If the decay width is very small (as it often might be because 
$N$ is quite large) then by continuity we expect that within our 
finite errors the correlators will behave just as they do for  
stable states. Conversely, if our correlator looks just like
that of a stable state, with an apparently well-defined energy plateau,
we can assume that the state is very narrow, and extract an energy.
This will certainly not always be the case. Sometimes we have accurate
correlators out to large $n_t$ where there is no sign of a plateau,
presumably because the state has a large decay width. We shall
perform a heuristic analysis of some of these cases when we come to them.
The interesting conclusion will be that this leads to an energy much higher 
than one would naively guess by looking at the effective energies.

We remark that the exact eigenstates of $H$ consist of asymptotic
states composed of any stable flux tubes and scattering states of 
these. (And in addition, at finite $N$, of stable glueballs.) 
In particular this includes scattering
states of fundamental and antifundamental flux tubes with various
relative momenta. However our basis of operators will usually
(although not intentionally) have a small overlap on these,
and so we usually will not see them in our calculation. We will
comment further on this when we consider examples of what
are presumably unstable states.

\begin{table}[t]
\begin{center}
\begin{tabular}{|c|c|c|c|}\hline
Rep & $a^2\sigma_R$ & $\sigma_R/\sigma_f$ & $C_R/C_f$ \\ \hline
 f      &   0.007365(7)    & --  &   -- \\
 2A   &   0.011980(30)  &  1.627(5)    &   1.6     \\
 2S   &   0.016536(70)   &  2.245(10)  &   2.286 \\
 3A  &   0.013571(50)   &  1.842(8)    &   1.8     \\
 3M  &  0.02101(14)     &   2.853(20)  &   2.829 \\
 3S   &  0.02799(21)     &   3.800(30)  &   3.857 \\
 adj  &  0.015072(75)   &   2.046(11)  &   2.057 \\
 84   &  0.020212(81)   &  2.744(12)  &    2.714 \\
 120 &  0.02458(22)     &   3.337(30)  &   3.4    \\ \hline
\end{tabular}
\caption{String tensions for various representations (see text) 
in SU(6) at $\beta=171$. Also ratios to the fundamental and
predictions of Casimir scaling.}
\label{table_sigma_R}
\end{center}
\end{table}

\section{Spectrum results}
\label{section_results}

In this section we present our results. Before entering into details we list 
in Table~\ref{table_sigma_R} the string tension $\sigma_R$  that we obtain
by fitting the absolute ground state energy, $E_0(l)$, for each representation 
with the Nambu-Goto expression in eqn(\ref{eqn_E0NG}) plus a $O(1/l^7)$ 
correction, i.e.
\begin{equation}
E_0(l)
=
\sigma l
\left\{1 - \frac{\pi}{3}\frac{1}{\sigma l^2}\right\}^{\frac{1}{2}}
+
\frac{c}{l^7}
\label{eqn_E0NG}
\end{equation}
We use this correction because it is the  leading correction to
the universal terms 
\cite{OAZK},
but since any correction will only affect $E_0(l)$ at small $l$ our
particular choice does not affect the value of the 
extracted string tension. We compare the ratio $\sigma_R/\sigma_f$
to the ratio of the quadratic Casimirs, $C_R/C_f$. There are old
arguments for such `Casimir scaling' (see 
\cite{BLMT_strings}
for a discussion and references)
as well as newer ones, e.g.
\cite{Nair}.
We see from Table~\ref{table_sigma_R} that it works remarkably well. 
This corroborates earlier studies
\cite{BLMTUW_imp,BBMT_kd3}
for some of these representations. (As well as older studies in SU(3)
of open flux tubes, e.g.
\cite{CSsu3}.)
The values of $E_0(l)$ that go into these fits are listed in
Tables~\ref{table_n6b171gs_a} and \ref{table_n6b171gs_b} where
we also show the lattice sizes used. For completeness we include 
the values for the fundamental representation obtained in our earlier work
\cite{AABBMT_fd3}.

\begin{table}[tb]
\begin{center}
\begin{tabular}{|cc|cccccc|}\hline
\multicolumn{8}{|c|}{ $aE(l)$ } \\ \hline
 $l$ & $l_\perp\times l_t$ &  f  & 2A  &  2S  &  3A &  3M &  3S  \\   \hline
16  & $100\times 200$  & 0.0777(3) &   0.1460(14)   &   0.2256(29) &
0.1742(11) & 0.2705(61) & 0.395(10)    \\ 
20  & $70\times 120$  & 0.1176(5) &   0.2088(17)     &  0.2955(57)   &
0.2433(21)   & 0.3723(99) &  0.529(10)   \\ 
24  & $48\times 60$   &  0.1528(9)  &  0.2649(23)   &  0.3669(42)   &
0.3020(32)    & 0.4593(80) & 0.651(12)     \\
28  & $48\times 60$   &  0.1842(8)  &  0.3198(29)    &  0.4490(53)  &
0.3569(39)    & 0.5720(86)  &  0.781(15)   \\
32  & $40\times 48$   &  0.2177(10) &  0.3633(22)  &  0.5067(68)   &
0.4198(53) & 0.6304(107) &  0.855(15)   \\
36  & $40\times 48$   & 0.2490(12)  &  0.4192(25)    &  0.5777(70)   &
0.4762(50)     & 0.7411(126) & 0.963(25)     \\
40  & $48\times 48$   &  0.2817(14) &   0.4615(42)   &  0.6504(82)   &
0.5259(67)    & 0.8173(123)  & 1.154(30)      \\
44  & $48\times 48$    &  0.3113(14)  &   0.5144(50)   &   0.7094(132)
&  0.5806(74)    & 0.9102(165) & 1.219(49)   \\
48  & $48\times 48$    &  0.3425(13) &   0.5624(40)  &   0.7818(96)
&  0.6405(79)    & 1.0101(197)  & --    \\
52  & $52\times 52$    &  0.3723(10)  &  0.6183(60)   &   0.8736(104)
& 0.7015(83)    & 1.1125(300) &  1.473(34)    \\  
64  & $64\times 64$     &  0.4637(17) &  0.7661(109)  & 1.0789(229)
& 0.8633(139)   & 1.3518(194)  & 1.837(73)     \\ \hline
\end{tabular}
\caption{The energies, $E(l)$, of the lightest flux tubes of length
  $l$ (all $l$ in lattice units) and $p=0$, and with the flux belonging to the
  indicated representations. The fundamental (f) values are from 
\cite{AABBMT_fd3}.
For SU(6) at $\beta=171.0$.}
\label{table_n6b171gs_a}
\end{center}
\end{table}

\begin{table}[tb]
\begin{center}
\begin{tabular}{|cc|ccc|}\hline
\multicolumn{5}{|c|}{ $aE(l)$ } \\ \hline
 $l$ & $l_\perp\times l_t$  &  $\underline{84}$ &  $\underline{120}$ &  adj  \\   \hline
16  & $100\times 200$  &   --           &  --        &   0.1658(66)  \\ 
20  & $70\times 120$  &  0.3658(35)  &  0.4460(47)  &   0.2568(70)  \\ 
24  & $48\times 60$   &   0.4476(55)  & 0.5576(85)   &   0.3327(42)   \\
28  & $48\times 60$   &   0.5462(53)  &  0.6725(148)  &   0.4071(57)  \\
32  & $40\times 48$   &   0.6297(58) &   0.7760(151)  &    0.4607(66)  \\
36  & $40\times 48$   &   0.7031(70)  &   0.8658(201)  &   0.5277(61)  \\
40  & $48\times 48$   &   0.8003(68) &   0.9625(243)   &   0.5955(89)  \\
44  & $48\times 48$   &   0.8950(112)  &  1.0866(329)   &   0.6561(119)  \\
48  & $48\times 48$   &   0.9674(163)  &    1.1767(326)  &   0.7291(139) \\
52  & $52\times 52$   &   1.0658(250)   &   --   &   0.7845(207) \\  
64  & $64\times 64$    &   --                &  --     & 0.954(38)    \\ \hline
\end{tabular}
\caption{The energies, $E(l)$, of the lightest flux tubes of length
  $l$ (all $l$ in lattice units) and $p=0$, and with the flux belonging to the
  indicated representations. For SU(6) at $\beta=171.0$.}
\label{table_n6b171gs_b}
\end{center}
\end{table}

\subsection{finite volume corrections}
\label{subsection_volume}

Calculations on $l\times l_\perp \times l_t$ lattices will suffer
finite volume corrections if $l_\perp$ and $l_t$ are not large enough.
This problem becomes more severe as $l$ decreases. Some checks have
been performed in
\cite{HMMT,AABBMT_k2d3}  
for $k=2$ flux tubes, and in
\cite{AABBMT_fd3}
for excited states of $k=1$ flux tubes. Since our calculations
are now more accurate, it is worth revisiting this question.

We focus on our shortest flux tube, where we employ a 
$16\times 100\times 200$ lattice. We are confident that $l_t/a=200$ is
long enough since $e^{-El_t} = e^{-200aE(l=16)}$ is negligible for all the
$l/a=16$ flux tube energies listed in  Tables~\ref{table_n6b171gs_a} 
and \ref{table_n6b171gs_b}. We therefore test whether 
$l_\perp/a=100$ is large enough and we do this by performing calculations
on   $16\times l_\perp/a\times 200$ lattices with $l_\perp/a = 20,40,60,80$.
To speed up these very slow calculations we use a much reduced basis of
operators - just the simplest Polyakov loops at various blocking levels. 
This still allows us to obtain accurate values for the ground states 
but not for any of the excited states. (Which is why we introduced our extended 
operator basis in the first place.) So for the excited states we continue 
to rely on the study in
\cite{AABBMT_fd3}
and the rescaling of those results to our lattice spacing.
 
In Table~\ref{table_n6b171ERl16} we show our results for the ground states in 
various representations. We see that the fundamental flux tube suffers
no finite volume corrections for $l_\perp/a \geq 40$ within the statistical 
uncertainty of about $\pm 1\%$. For the higher representations there are
still visible corrections for  $l_\perp/a = 40$ but $l_\perp/a \geq 60$ appears to 
be safe at the $\pm$2 or 3 percent level of our statistical errors.
It thus appears that $l_\perp/a = 100$ is in fact a very safe and conservative 
choice. This provides further evidence that the energies calculated in this paper
are not afflicted by significant finite size corrections.

\begin{table}[tb]
\begin{center}
\begin{tabular}{|c|ccccc|}\hline
\multicolumn{6}{|c|}{ $aE_R(l=16;l_\perp)$ } \\ \hline
 $R$  & $l_\perp=20$ & $l_\perp=40$  & $l_\perp=60$   &  $l_\perp=80$  &   $l_\perp=100$  \\   \hline
 f (k=1)     &   0.0742(10)    & 0.0781(8)     & 0.0781(11)    & 0.0766(13)    &  0.0777(3)     \\
 k=2A   &   0.1167(18)    & 0.1385(18)   &  0.1430(28)   & 0.1460(20)    &  0.1460(14)    \\
 k=2S    &  0.2335(18)    &  0.2243(19)  &  0.2260(31)   &  0.2280(24)   &  0.2256(29)    \\
 k=3A    &  0.1292(32)    & 0.1624(26)   &  0.1706(35)   &  0.1748(26)   &  0.1742(11)     \\
 k=3M   &   0.2521(48)   &  0.2573(40)  & 0.2675(64)     & 0.2699(60)    &  0.2705(61)    \\
 k=3S    &   0.4148(41)   &  0.364(12)    &  0.390(9)       & 0.421(5)   &  0.409(4)    \\
  Adj (k=0)    &    0.1553(44)   &  0.1652(30) &  0.1692(47)   &  0.1796(44)  &  0.1658(66)  \\ \hline
\end{tabular}
\caption{The energy, $E_R(l)$, of the lightest flux tube of length
  $l=16$ (all $l$ in lattice units) on a $16\times l_\perp\times 200$ lattices,
 and with the flux belonging to the representation $R$.
For SU(6) at $\beta=171.0$.}
\label{table_n6b171ERl16}
\end{center}
\end{table}

In Table~\ref{table_n6b171ERl40} we again show some results for the ground 
states (and also for some excited states) in various representations, 
but this time for a much longer flux tube, 
$l/a=40$. This confirms that a transverse size $l_\perp/a=32$ is already large enough,
and the sizes we have actually used are very conservative.

\begin{table}[tb]
\begin{center}
\begin{tabular}{|c|ccc|}\hline
\multicolumn{4}{|c|}{ $aE_R(l=40;l_\perp)$ } \\ \hline
 $R$  & $l_\perp=16$ & $l_\perp=32$  & $l_\perp=48$   \\   \hline
 f (k=1)  &  0.2790(24) &  0.2802(28) &  0.2817(14)      \\
  f (k=1)$\star$  &   0.527(10)   &  0.522(7)   &  0.507(3)      \\
 k=2A   &   0.438(4)    & 0.465(7)      &  0.462(5)     \\
 k=2A$\star$   &   0.718(4)    & 0.663(11)   &  0.655(11)     \\
 k=2S    &  0.673(5)    &  0.661(7)     &  0.650(9)    \\
 k=3A    &  0.484(9)    & 0.530(8)      &  0.526(7)      \\
 k=3A$\star$   &  0.774(18)    &  0.719((10)   &  0.732(15)      \\
 k=3M   &   0.814(7)   &  0.799(24)    &  0.817(13)     \\
 k=3S    &   1.14(2)     &  1.116(16)    &  1.154(29)         \\
  Adj (k=0)  &   0.566(6)   &  0.584(10)    &  0.560(9)       \\ \hline
\end{tabular}
\caption{The energy, $E_R(l)$, of the lightest flux tube of length
  $l=40$ (all $l$ in lattice units) on a $40\times l_\perp$ spatial volume,
 and with the flux belonging to the representation $R$.
For SU(6) at $\beta=171.0$.}
\label{table_n6b171ERl40}
\end{center}
\end{table}

Finite size corrections also affect the screening of one representation to another,
as shown in Tables 2,3 of
\cite{AABBMT_k2d3}.
This is relevant because it is only when the screening is very weak that
we can categorise the states as being (almost entirely) in $k=2A$ and $k=2S$
rather than just $k=2$ (and similarly for our other representations).
We therefore perform a similar analysis here. We define the normalised overlap
\begin{equation}
O_{2AS}(b) = \frac{\langle \Phi^\dagger_{2A,b}(t=0)\Phi_{2S,b}(t=0)\rangle }
{\langle \Phi^\dagger_{2A,b}(t=0)\Phi_{2A,b}(t=0)\rangle^{1/2}          
\langle \Phi^\dagger_{2S,b}(t=0)\Phi_{2S,b}(t=0)\rangle^{1/2} }
\label{eqn_2ASover}
\end{equation}
where $\Phi_R(t)$ is the simple Polyakov loop at blocking level $b$ 
and representation
$R$, in the time-slice $t=0$ and, as usual, summed over spatial sites so as 
to have zero transverse momentum. (Obviously we will average over all equal times.)
The range of values of $b$ is restricted by the fact that a `blocked link'
\cite{MT_block,AABBMT_fd3} 
joins lattice sites that are separated by $2^{b-1}$ lattice sites. 
So for $l/a=16$ it only
makes sense to consider $1\leq b \leq 5$. Essentially, loops at 
blocking level $b$ are smeared over distances significantly 
greater than this separation  $2^{b-1}$. Thus the highest blocking
level shown typically involves operators that overlap over the 
boundary of the torus and these can be affected by strong finite volume 
corrections. 

Bearing the above in mind, we show our results for the overlap $O_{2AS}(b)$ 
in Table~\ref{table_2A2Sover}. We remark that the calculations with $l/a\neq 16$
are mostly with lower statistics, designed to be sufficient for our purposes here. 
We also calculate Polyakov loops in the (usually) longer $y$ direction, and
this gives us some values of  $O_{2AS}(b)$ at small $l_\perp$ (now $=l_x$) and
larger $l$ (now $=l_y$) which we also present in Table~\ref{table_2A2Sover}.
We conclude from this Table that: \\
1) for very small $l$ the overlap  $|O_{2AS}(b)|$ is large for all $b$; \\
2) and for fixed $l$ the values of $|O_{2AS}(b)|$ grow as $l_\perp$ decreases; \\
3) but  $|O_{2AS}(b)|$ rapidly decreases to values consistent with zero as $l \to \infty$,
and this is so for any fixed $b$ and appears to be the case for any fixed $l_\perp$ as well.

We conclude that for long flux tubes on large volumes, we can safely ignore screening
and label states as $k=2A$ and $k=2S$. Indeed we see that it is only when $l$ or $l_\perp$
are close to the phase transition at $l_c$ that screening is significant.
Our results for $k=3A,3M,3S$ are very similar and the vacuum expectation value of the
adjoint loop shows very similar trends. In practice this means that in 
Tables~\ref{table_n6b171gs_a},~\ref{table_n6b171gs_b} it is only for $l/a=16$ (and
 $l/a=20$ for some $k=3$) that the states have needed to be extracted using 
the whole $k=2$
or $k=3$ basis (and we have then assigned the $A,M,S$ labels on the basis of 
what component dominates the wave function).

\begin{table}[tb]
\begin{center}
\begin{tabular}{|cc|llllll|}\hline
\multicolumn{8}{|c|}{ 2A/2S overlap } \\ \hline
 $l$ & $l_\perp\times l_t$  &  bl=1  & bl=2  & bl=3  &  bl=4  & bl=5   & bl=6  \\  \hline
13  & $60\times 200$  & 0.292(48) & 0.344(53) & 0.384(56) & 0.438(58) &  --  & --  \\
14  & $60\times 200$  & 0.122(25) & 0.157(31) & 0.187(36) & 0.227(42) &  --  & --  \\  \hline
16  & $20\times 200$  & 0.172(6) & 0.234(7) & 0.285(8) & 0.356(9) & 0.491(8) & -- \\
16  & $40\times 200$  & 0.053(3) & 0.076(4) & 0.097(5) & 0.129(6) & 0.206(8) & -- \\
16  & $60\times 200$  & 0.036(3) & 0.053(4) & 0.067(5) & 0.088(6) & 0.136(8) & -- \\
16  & $80\times 200$  & 0.034(2) & 0.048(3) & 0.061(4) & 0.081(5) & 0.122(6) & -- \\
16  & $100\times 200$ & 0.032(3) & 0.047(3) & 0.059(4) & 0.076(5) & 0.116(7) & -- \\   \hline
20  & $16\times 200$  & 0.071(2)  & 0.125(3) & 0.175(3) & 0.259(4) & 0.416(3) & --   \\
40  & $16\times 200$  & 0.001(1)  & 0.002(1) & 0.009(1) & 0.035(2) & 0.189(2) & --  \\
60  & $16\times 200$  & 0.000(1)  & 0.001(1) & 0.000(1) & 0.006(1) & 0.090(2) & -- \\
80  & $16\times 200$  & 0.001(1) & 0.001(1)  & 0.001(1) & 0.001(1) & 0.041(2) & --  \\
100 & $16\times 200$  & 0.000(1) & 0.001(1)  & 0.001(1) & 0.000(1) & 0.020(1) & -- \\ \hline
20  & $70\times 120$  & 0.010(5) & 0.015(8) & 0.019(12)& 0.028(16)& 0.037(19) & -- \\
24  & $48\times 60$   & 0.002(4) & 0.001(5) & 0.003(7) & 0.011(9) & 0.022(13) & -- \\
32  & $40\times 48$   & 0.001(4) & 0.003(3) & 0.000(6) & 0.003(7) & 0.001(11) & 0.107(17) \\
48  & $48\times 48$   & 0.001(3) & 0.001(3) & 0.003(5) & 0.001(6) & 0.004(5)  & 0.011(5)  \\  \hline
\end{tabular}
\caption{The modulus of the normalised overlaps $|O_{\mathrm{2AS}}(\mathrm{bl})|$ 
of blocked Polyakov loops in
the 2A and 2S representations, for blocking levels bl, as defined in eqn(\ref{eqn_2ASover}).
On lattices of various sizes (shown in lattice units).}
\label{table_2A2Sover}
\end{center}
\end{table}

\subsection{k=2A, 2S}
\label{subsection_k2}

In the $k=2$ sector we focus on the irreducible representations in 
$f \otimes f$, i.e. the totally antisymmetric, $2A$ and the totally 
symmetric, $2S$
\cite{BLMT_strings}. 
The $k=2$  sector contains other representations, e.g. from the decomposition 
of $f \otimes f \otimes f \otimes {\overline{f}}$, but one expects these to 
have higher energies, and we do not consider them here. As we have remarked
above, the dynamics appears to respect these representations very well, despite
the potential mixing from gluons in the vacuum. Only for $l\sim l_c$ is
there significant mixing.

The lightest $k=2$ flux tube is essentially pure $k=2A$. We see from 
Table~\ref{table_n6b171gs_a} that it is lighter than two fundamental flux tubes
(which would also be $k=2$) so this flux tube is absolutely stable.
Its calculation therefore provides a `benchmark' for what constitutes a `good' 
energy calculation in this paper. The energy is calculated from the correlator
$C(t=an_t)$ of our variationally selected best trial wave-functional for the state.
We can define an effective energy by
\begin{equation}
aE_{eff}(n_t) = -\ln\frac{C(n_t)}{C(n_t-1)}
\label{eqn_Eeff}
\end{equation}
and note that if $C(t)$ is independent of $t$ for $t\geq t_0$ (within errors), then 
this implies that it is given by a single exponential, $C(n_t)/C(0) = |c|^2 e^{-aEn_t}$
for $t\geq t_0$ (within errors). So to calculate $aE$ we need to identify a
plateau in the values of $aE_{eff}(n_t)$ and the quality of our calculation is reflected in
how convincing this plateau is.

In Fig.~\ref{fig_Eeffk2Aq0_n6f} we plot our values of $aE_{eff}(n_t)$ for various values of $l$.
We also show our final energy estimate in each case by the horizontal lines. We have excluded
values at larger $n_t$, once the errors have become larger than $\sim 15-20\%$ since these carry
little information and merely clutter the plot. (In addition, at large $n_t$ the correlations 
within the Monte Carlo sequence become very long and our error estimates become increasingly 
unreliable.) We can see that we have a well-defined energy plateau for all our values of $l$,
although the length of the plateau shortens as $l\uparrow$ since $e^{-E_nt}$ will disappear
into the statistical noise more quickly with increasing $t$ for larger $E_n$.

We fit these energies with the Nambu-Goto formula in eqn(\ref{eqn_E0NG}), together with a 
theoretically motivated $O(1/l^7)$ correction, which however plays no significant role
in the fit. We extract the string tension $a^2\sigma_{2A}$ and plot 
in Fig~\ref{fig_EgsQallk2a_n6f}
the values of $E_0(l)$ versus $l$, with both expressed in units of the string tension.
We see a very clear near-linear increase characteristic of linear confinement. We also see
that the pure Nambu-Goto prediction appears to fit very well. 

Uniquely for the absolute ground state the expansion of the Nambu-Goto prediction 
for the energy  $E_0(l)$ in powers 
of $1/l^2\sigma$ converges right through the range of $l$ where we have calculations;
indeed all the way down to  $l\surd\sigma = \pi/3 \sim 1.1 < l_c\surd\sigma$. This
provides an opportunity to test not just the resummed Nambu-Goto expression, but the 
individual power correction terms predicted to be universal
\cite{OAZK}. 
To do this we normalise  $E_0(l)$ to the leading $\sigma l$ piece, so that we can 
readily expand the scale, and compare to various `models' for $E_0(l)$. 
This produces Fig.~\ref{fig_DENGgsq0P+k2a_n6f}. Here we see that the free string expression 
is good all the way down to $l\surd\sigma_{2A} \sim 2$ which is close to the deconfining length, 
$l_c$, indicated by the vertical red line. And we note that a  $O(1/l^7)$ correction can
describe the deviations from Nambu-Goto for $l\surd\sigma_{2A} \leq 2$. However we also
see that including just the leading universal correction, i.e. $E_0(l) = \sigma l -\pi/6l$
\cite{LSW,PS}, 
is indistinguishable from Nambu-Goto within the errors in the range of $l$ where
the latter well describes $E_0(l)$. However if we only include a linear 
$\sigma l$ piece, then this does not fit at all. Thus we have a quite accurate
confirmation of the presence of the   universal L\"uscher correction, but not really 
much more than that. The reason for this is that the universal corrections to $E_0(l)$
have small coefficients, since they represent just the zero-point energies of the
string fluctuation modes, which indeed is why the expansion converges down to
small $l$. (One can do better with the fundamental flux tube
\cite{AABBMT_fd3}, 
since $\sigma$ is smaller there, and it is in that case that one may realistically hope
to pin down all the universal corrections.) 

It is worth quantifying how well we  can constrain the L\"uscher
correction with the k=2A ground state. We find 
\begin{equation}
aE_{2A}(l) =\sigma_{2A} l -c_{eff} \frac{\pi}{6l} \quad ; \quad c_{eff} =
1.05(15) \ \mathrm{for} \ l \surd\sigma_{2A} \geq 2.5,
\label{eqn_corML2A}
\end{equation}
which is a usefully accurate test of this universal coefficient.

We now turn to states with non-zero longitudinal momenta. In
Fig.~\ref{fig_EgsQallk2a_n6f} we also plot the ground state energies for the 
lowest two non-zero momenta along the $l$-torus, $p=2\pi /l$ and $p= 4\pi /l$. 
We find that there is a unique such 
state for  $p=2\pi /l$ and it has $P=-$. For $p= 4\pi /l$ we find two apparently
degenerate ground states, one with $P=+$ and one with $P=-$. All this is just as
expected from Nambu Goto where the  $p=2\pi /l$ state has one phonon, and hence $P=-$,
and  the  $p=4\pi /l$ ground states have either one phonon carrying the whole momentum, 
with $P=-$, or two phonons sharing the momentum, and hence $P=+$. We also show
in Fig.~\ref{fig_EgsQallk2a_n6f} the ground state energy of two (non-interacting)
fundamental flux tubes of length $l$ carrying the same total momentum. We see that
this state always has a higher energy than that of the corresponding $k=2A$ flux tube
showing that the latter is indeed stable. 

Since the only parameter in Nambu-Goto is the string tension, which is obtained by
fitting the  $p=0$ state, the  Nambu-Goto predictions shown for  $p=2\pi /l$ and $p= 4\pi /l$ 
have no free parameters. It is therefore remarkable that the agreement is so precise and 
extends to our smallest values of $l$. Of course some of the energy comes from $p^2$ and 
so it is useful to perform a comparison with this subtracted. We therefore define the
quantity:
\begin{equation}
\Delta E^2(q,l)
=
E^2(q;l) - E^{NG \ 2}_{0}(l) 
- \left ( \frac{2\pi q}{l}\right )^2
\stackrel{NG}{=} 4\pi\sigma (N_L+N_R),
\label{eqn_exq} 
\end{equation} 
using eqns(\ref{eqn_EnNG}) and (\ref{eqn_E0NG}). This exposes the excitation energy
predicted by Nambu-Goto. We plot the ratio $\Delta E^2(q,l)/ 4\pi\sigma$  in 
Fig.~\ref{fig_DE2NGgsq0012k2a_n6f}. We see that the integer-valued contribution
of the excitation energy is very accurately confirmed for all $l$, even for very short 
flux tubes which certainly do not `look like' thin strings. This is something that
we have already observed for fundamental flux tubes
\cite{AABBMT_fd3}
but here we know that the flux tube is a bound state with, therefore, some extra
internal structure. From the comparison in Fig.~\ref{fig_EgsQallk2a_n6f} between
the $k=2A$ energy and that of two free $k=1$ flux tubes, we infer that the binding energy
is not very large, so that at small $l$ the $k=2A$ flux tube will be a `blob' rather
than a `thin string'. It is therefore remarkable that its excitation spectrum
should be so precisely that of a free thin string. 

In Fig.~\ref{fig_DE2NGgsq0012k2a_n6f} we also show what happens if one
excludes the zero-point energy from the Nambu-Goto formula. We see a very visible
shift for both $p=0$ and $p=2\pi/l$. (It would be  pointless to go to higher $p$ 
since the errors are too large there.) For $p=0$ this is just another presentation
of our result in eqn(\ref{eqn_corML2A}), however it is interesting to see that
the $p\neq 0$ spectrum also reveals the presence of this zero-point energy.
We do not quantify it further because it would add little to  eqn(\ref{eqn_corML2A}).

To assess the significance of these results for $p\neq 0$ it is worth stepping back 
and asking what we might expect if we make no assumption at all about the relevance
of stringy fluctuations. We would expect on general grounds that the absolute ground 
state of the flux tube is intrinsically translation invariant in the direction of the
flux tube, so can only have $p=0$. Thus the non-zero $p$ has to be carried by some 
additional excitation. Let us suppose that this is some particle of mass $m$. Then 
neglecting any interaction between this particle and the flux tube, the energy
of the combined system is
\begin{equation}
E(l;p) = E_{gs}(l) + (m^2+p^2)^{1/2}  \qquad ; \quad p=\frac{2\pi q}{l} 
\label{eqn_Epmodel} 
\end{equation} 
where $E_{gs}(l)$ is the (observed) energy of the absolute ground state. To 
decide whether this model has any plausibility, we plot  $E(l;p)$  for the
massless case, $m=0$, in Fig.~\ref{fig_EgsQallk2a_n6f} as the dashed lines.
We see that these are very close to the Nambu-goto predictions and
could provide a good first approximation to the observed spectrum.
It is therefore interesting to ask how this constrains the value of $m$. So
we calculate $m$ using eqn(\ref{eqn_Epmodel}) at each value of $l$
for $p=2\pi/l$, since these $p\neq 0$ energies are the most accurate,
and average the results for $l\geq l_0$, for various choices of $l_0$.
The result, in units of the string tension, is shown in 
Fig.~\ref{fig_muq1k2Ak3A_n6f}. (We also show the similar result of a 
similar analysis applied to flux tubes in the $k=3A$ representation.) 
Roughly speaking this tells us that $m^2/\sigma_{2a} \lesssim 0.1(1)$. This is to 
be compared to the known value of the mass gap in the SU(6) gauge theory
\cite{BLMT_d3,MT_98}
which is $m^2_G/\sigma_{2a} \sim 13$. Thus this `particle' cannot be an
excitation in the bulk space-time, and must be an excitation that
lives on the flux tube. In that case the obvious candidate is a 
massless stringy mode of the kind described by the Nambu-Goto free
string model.  Note that this of course means that the relationship 
in eqn(\ref{eqn_Epmodel}) is not the correct one. Note also that
although eqn(\ref{eqn_Epmodel}) is, numerically, very close to
eqn(\ref{eqn_EnNG}) for states where the massless phonons are either all
right or all left movers, this is no longer the case when both
right and left movers are present, e.g. the first excited $p=0$
state. As it happens, we shall shortly see that, although this state 
is badly described by the extension of eqn(\ref{eqn_Epmodel}),
it is also badly described by Nambu-Goto. However in the case of 
fundamental flux tubes, studied in
\cite{AABBMT_fd3},
one finds that Nambu-Goto works well for $l$ not very small, and 
thus eqn(\ref{eqn_Epmodel}) would be strongly disfavoured. In
addition a state with a ground state $p=0$ flux tube and an
additional particle would not couple to our operators as $N\to\infty$
in contrast to what one observes for the states with $p\neq 0$.
Our purpose in considering this simple model was to establish, in a
pedestrian way, that one must look to massless modes living on the flux tube
for the origin of the observed spectrum.

We turn now to the spectrum of excited states with $p=0$. We plot,
in Fig~\ref{fig_Eq0k2a_n6f}, the four lightest $P=+$ states, and the
two lightest  $P=-$ ones, as well the predictions of
Nambu-Goto for the lowest few energy levels. (We also plot 
some higher excitations for $l=32a$ and $l=52a$, which we shall return 
to shortly.) In Nambu-Goto the ground state, with no phonons, is
non-degenerate, with $P=+$, as is the first excited energy level which
has one left and one right moving phonon with momenta 
$p=\pm 2\pi/l$. The next energy level has four degenerate states
with the left and right moving phonons sharing twice the minimum momentum.
Since this can be carried by one or two phonons, two of these states have 
$P=+$ and two have  $P=-$. If the $2A$  flux tube states were close to Nambu-Goto, 
as they turn out to be for the case of  fundamental flux, we should find
our calculated energies clustering closely about the lowest three Nambu-Goto
energy levels. While we do indeed observe in  Fig~\ref{fig_Eq0k2a_n6f} that
the lightest two states do have parity  $P=+$, and the next two
$P=+$ states are roughly in the same energy range as the lightest two
$P=-$ states, we see nothing like the (near)degeneracy predicted by Nambu-Goto.
There is some evidence that the first excited $P=+$  state and the
lightest $P=-$ state approach the appropriate Nambu-Goto levels, 
and that  the second lightest $P=-$ state agrees with the Nambu-Goto
prediction for all but the smallest values of $l$.  However the observed
excited states are , in general, far from showing the Nambu-Goto degeneracies 
and are far from the Nambu-Goto predicted energies,
even for the largest values of $l\surd\sigma$. While the first excited state appears
to clearly approach the string prediction, even here it would be useful
to have some further evidence that it is asymptoting to that curve and not just
crossing it. It is useful to recall that for the fundamental flux tube
\cite{AABBMT_fd3},
the convergence to Nambu-Goto was rapid and unambiguous (albeit not as
rapid as for the ground state). The messiness of the picture in 
Fig~\ref{fig_Eq0k2a_n6f} is of course what one would have naively expected
for such a bound state flux tube, and the real surprise is the precise
stringy behaviour we have observed for the lightest states with non-zero 
momenta. One significant difference with the latter is that here the states are 
generally well above the threshold for decay. The lightest asymptotic decay products 
will be two fundamental flux tubes with equal and opposite transverse momentum. 
The energy of the threshold, corresponding to zero relative momentum, is shown
in  Fig~\ref{fig_Eq0k2a_n6f} and one is tempted to note that the deviation of
the first excited state from Nambu-Goto decreases as the phase space for decay 
decreases. We also show the energy of a decay state composed of a glueball
and a ground state $k=2A$ flux tube. We see that this is quite high and, 
in addition, such a decay will be  large-$N$ suppressed.  

To provide some more context for these states, we have also shown 
in Fig~\ref{fig_Eq0k2a_n6f}  the next 6 $P=+$ and 5 $P=-$ states 
for $l=32a$ and $l=52a$ (slightly shifted in $l$ for clarity).
The number of states is motivated by the fact that the next Nambu-Goto
energy level has 5 $P=+$ and 4 $P=-$ degenerate states, so 
we are also including at least one state, for each $P$, that will
approach a yet higher energy level as $l\to\infty$. (But note that the
extraction of the energies can be ambiguous for these massive
states.) The main message, considering all the $l=32a, 52a$ states, is
that there is no visible clustering in the energy of the states that
might suggest that they are converging to the Nambu-Goto energy levels,
except for the absolute ground state and perhaps the first excited
state, both of which are $P=+$ at our largest value of $l$ where
a clear gap has opened between them and the $P=-$ states -- 
as expected in Nambu-Goto.
For the first excited state there is a residual ambiguity: is it
the first excited state at lower $l$ that asymptotes to the Nambu-Goto level 
as $l\to\infty$, or is it perhaps the second, with the first 
`crossing' that level somewhere between $l=52a$ and $l=64a$ ? 
In fact our analysis in Section~\ref{section_massiveorstringy}
will address and resolve this issue. What we see in
Fig~\ref{fig_Eq0k2a_n6f},  particularly for $l=52a$, is 
very much  a continuous distribution of excited states
without any obvious level structure. This makes it hard, for example, to know
whether the near-coincidence of the second $P=-$ energy with the
Nambu-Goto prediction is in fact significant, or merely the chance
result of this near-continuous distribution of states.
What is clear from Fig~\ref{fig_Eq0k2a_n6f} is that we are very
far from the values of $l$ where the Nambu-Goto spectrum 
might become a good first approximation for these states.

The fact that these excited $p=0$ states are generally well above their decay 
thresholds raises some questions. The most important is how confident can we
be that we have extracted their `energies'? If the decay width is very small,
the propagator should have a pole in the complex energy plane very close to the 
real axis, and we would expect correlators designed for stable states to behave 
just as they do for a stable state, within the finite statistical errors, 
i.e. we should see an effective energy plateau that is lost in the statistical 
errors at larger $t$ before deviations from the plateau become visible.
We show the effective energy plateaux for the first excited state in 
Fig.~\ref{fig_Eeffk2Aq0ex_n6f}. We see that for large $l$ these plateaux are 
unambiguous and not so different from those of our stable ground state in 
Fig.~\ref{fig_Eeffk2Aq0_n6f}. As $l$ decreases, however, the apparent plateau 
shifts to larger $t$ and becomes increasingly ambiguous. This is very
different to what we observe in Fig.~\ref{fig_Eeffk2Aq0_n6f}. The likely reason 
for this is that the phase space for the decay of the first excited state grows 
as $l$ decreases (as we scan infer from Fig~\ref{fig_Eq0k2a_n6f}), and so
presumably does the decay rate. So it is interesting to perform a different 
analysis, at the smallest values of $l$, that attempts to take this finite
decay width into account. The relevant asymptotic states in this energy range
are those composed of two (unexcited) fundamental flux tubes with equal and opposite
transverse momenta. (Flux tubes with longitudinal momenta have larger energies.)
Obviously if we performed a variational calculation with a complete basis of $k=2$
operators, then these are the states we would obtain. However the operators we 
actually use are all of the form 
$\mathrm{Tr}_{2A}\{l_p\} \propto \mathrm{Tr}_f\{l_p\}^2 - \mathrm{Tr}_f\{l_p^2\}$ with $l_p$ some winding
operator. The  $\mathrm{Tr}_f\{l_p\}^2$ piece represents two fundamental flux tubes at
zero spatial separation, which can be re-expressed as a sum over all relative 
momenta. However the projection onto any such state with given momentum will be very 
small, so a variational calculation performed at $t=a$, as ours is, will not pick out
these states. However the overall projection onto all these states is not small,
and a heuristic procedure is to  perform a fit to the correlation function
that is in terms of these asymptotic scattering states, but with an
amplitude that encodes a slightly unstable state. We choose, again heuristically,
a Breit-Wigner form. So we fit to:
\begin{equation}
C(t)
=
\sum_{\vec{p}\neq 0} |c_{BW}|^2 e^{-E(p)t}
\quad ; \quad E(p) = 2E_f(p) , \ 
|c_{BW}|^2 = \frac{c}{(E-E_0)^2+(E_0\Gamma_0)^2}
\label{eqn_corBW}
\end{equation} 
where $E_f$ is the lightest energy of a fundamental flux tube with transverse 
momentum $p$, $c$ is a constant fixed by normalisation, $E_0$ is the real part 
of the pole energy and $\Gamma_0$ the (full) width. We either
sum over a discretisation of the momentum integral, or use the transverse
momenta dictated by the size of the transverse torus. (In practice it does not
matter which we use.) In Fig~\ref{fig_Eeffk2Aq0ex_l16_n6f} we display the values
of $aE_{eff}$ for $l=16a$, on a blown-up scale, and display different fits.
The red line arises from a conventional fit with an excited state in addition
to the desired lightest state. Here the lightest state is at 
$aE=0.25$, and the heavier one is at $aE=0.49$, with relative probabilities
$15\%$ and $85\%$ respectively. With such a low overlap, we can have little
confidence in the robustness of this lightest state. The alternative fit based on
eqn(\ref{eqn_corBW}) is shown by the solid black line and corresponds
to  $aE_0=0.475$ and $\Gamma_0=0.065$. (The dotted black line corresponds
to a sum over scattering states with uniform probability.) We see that the
value of the energy is close to but larger than $E_{eff}(t=a)$ whereas a search
for a large-$t$ plateau leads (as in our first fit above) to a much lower result. 
This is characteristic of such fits. We note 
that it is no coincidence that in our first, conventional, fit the
dominating `excited' state at $aE=0.49$ is close to our Breit-Wigner pole 
in the alternative fit based on eqn(~\ref{eqn_corBW}). This gives us 
confidence that this is most likely the actual energy of this unstable
excited state. We note that applying such a procedure would raise
the energy estimate significantly closer to the Nambu-Goto prediction.
For example, a similar analysis 
at $l=20a$ would give $aE_p=0.485$, with  $\Gamma_p=0.050$,
rather than the value $aE \simeq 0.38$ from a plateau estimate, and this would
approximately halve the discrepancy with Nambu-Goto. The effect is even more 
marked at $l=16a$. Clearly what we need is sufficient statistical accuracy
to distinguish between the two different $n_t\to\infty$ values of $aE_{eff}(n_t)$
in  Fig~\ref{fig_Eeffk2Aq0ex_l16_n6f}. Moreover it would be useful to see the
stability of such an analysis to the presence of a second heavier excited state
(which surely contributes at some level). Nonetheless, while we cannot be definitive
on this, it is plausible that where the apparent plateau is indistinct because
it is at large $n_t$, and in addition the state has a large phase space to decay, 
the actual energy of the `resonant' flux tube is much closer to the value of 
$aE_{eff}(t\to 0)$ than to   $aE_{eff}(t\to \infty)$. In the present
case this would
suggest values for the first excited state that are closer to the Nambu-Goto
prediction at small $l$ than our conventional estimates shown in Fig~\ref{fig_Eq0k2a_n6f}.
So it is not possible for us to be certain how much of the large apparent 
deviation from Nambu-Goto is due to the extra modes associated with the internal 
structure of the $k=2A$ flux tube, and how much is a consequence of the fact that
these states are unstable.

Two remarks. The first is that none of the above caveats apply to the ground states 
with $p\neq 0$ shown in Fig.~\ref{fig_EgsQallk2a_n6f}. Here the effective 
energy plateaux (which we do not show) 
start at small $t$ and typically become increasingly well-defined as $l$ 
decreases. The second remark is that one might wonder if some of the apparent
downward drift in $E_{eff}(n_t)$ that we see at large $n_t$ in  
Fig.~\ref{fig_Eeffk2Aq0ex_n6f} is not due some small admixture of the
ground state in our variationally estimated excited state wave function.
Since our variational ground state wave function has a typical 
overlap onto the ground state of $\sim 0.985(15)$ (which can be inferred
from the $E_{eff}(n_t)$ values shown in Fig.~\ref{fig_Eeffk2Aq0_n6f})
we can estimate the maximum such contribution to the excited  $E_{eff}(n_t)$
in Fig.~\ref{fig_Eeffk2Aq0ex_n6f}, and it turns out to be invisible 
for $l/a\geq 24$ (at our level of accuracy)
and only possibly becomes visible for $n_t\geq 15$
for $l/a=16,20$. That is to say, it is essentially irrelevant here.

We turn now to flux tube states obtained by performing calculations
with operators projected onto the $k=2S$ representation, i.e
$\mathrm{Tr}_{2S}\{l_p\} \propto \mathrm{Tr}_f\{l_p\}^2 + \mathrm{Tr}_f\{l^2_p\}$ with $l_p$ some winding
operator. (For the $l=16a$ $p=0$ ground state we obtain a cleaner
variational state by using the full $2A \oplus 2S$ basis, and 
that is what we show here. The admixture of $2A$ is small and 
so it still makes sense to label the state as $2S$, as we do.)
We know that these will be heavier than the corresponding $k=2A$ states
\cite{BLMT_strings,BBMT_kd3}
and so we expect all of them to be unstable  as well as having larger
statistical errors. In Fig.~\ref{fig_EgsQallk2s_n6f} we show the ground states
with the lowest longitudinal momenta. We also show the energies of the lightest 
decay products in each case. Just as for $k=2A$ the energies are remarkably close to
the Nambu-Goto predictions, as emphasised by comparing the actual excitation
energies in Fig.~\ref{fig_DE2NGgsq0012k2s_n6f}. In Fig.~\ref{fig_Eeffk2Sq0_n6f}
we show the effective energies for the absolute $p=0$ ground state. We also
indicate the expected decay thresholds on the right side of the figure. It seems clear
that $E_{eff}(n_t)$ does possess extended plateaux very different from the decay 
thresholds in the  lowest $l$ cases where we have accurate results to large $t$.
So while the quality of the calculations is markedly inferior to the $k=2A$
case, we have confidence in our extraction of the energies plotted in
Fig.~\ref{fig_EgsQallk2s_n6f}. The situation with the excited $p=0$ states is
however much worse and we are unable to extract the corresponding energies. Our
problem is illustrated by Fig.~\ref{fig_Eeffk2Sq0ex_n6f} where we plot
the effective energies for the `state' selected by our variational procedure 
as the first excited $p=0$ state. We cannot identify a plausible energy plateau
for any value of $l$, and $E_{eff}(n_t)$ is consistent with a decrease towards the
decay thresholds shown. In Fig.~\ref{fig_Eeffk2Sq0ex_l16_n6f} we repeat the 
exercise in Fig~\ref{fig_Eeffk2Aq0ex_l16_n6f}, now for the $l/a=16$ $k=2S$ flux tube.
The fit using eqn(\ref{eqn_corBW}) works very well, and corresponds to
an energy $E_0=0.58$ and a width $\Gamma_0=0.1$. The two exponential fit is
less convincing and corresponds to energies $0.205$ and $0.595$ with overlaps
squared of $0.1$ and $0.9$ respectively. This begins to point rather unambiguously
to an energy estimate of $E \sim 0.58$ and hence $E/\surd\sigma_{2S} \sim 4.6$
at $l\surd\sigma_{2S} \sim 2.1$. We note that this is below, but not far below,
the Nambu-Goto prediction. A similar conclusion follows for $l/a=20$. It is thus
plausible that this unstable first excited state is indeed quite close to Nambu-Goto
although this would be far from apparent using a conventional analysis.

\subsection{k=3A, 3M, 3S}
\label{subsection_k3}

In the $k=3$ sector we focus on the irreducible representations in
$f \otimes f \otimes f$, which are the totally antisymmetric, $3A$,
the mixed, $3M$, and the totally symmetric, $3S$. We know from
earlier work 
\cite{BLMT_strings,BBMT_kd3}
that the corresponding string tensions are very close to the
predictions of Casimir scaling (see also Table~\ref{table_sigma_R})
and so, as we shall see, the ground $3A$ states are stable, the
$3M$ states nearly so, and the $3S$ states are highly unstable.

In Fig.~\ref{fig_EQgsk3A_n6f} we plot the lightest energies of $k=3A$ flux 
tubes with longitudinal momenta $p=0, 2\pi/l, 4\pi/l$. Just as for the 
corresponding $k=2$ flux tubes, we see excellent agreement with Nambu-Goto
all the way down to $l\sim l_c$. The relevant asymptotic  decay states are 
not just 3 fundamental flux tubes, but also a stable $k=2A$ flux tube
with a fundamental one. The latter is lighter and the thresholds for both
are plotted as the black lines in Fig.~\ref{fig_EQgsk3A_n6f}, demonstrating
the stability of the $k=3A$ states. As we see in Fig.~\ref{fig_Eeffk3Aq0_n6f},
for the absolute ground state, we have very well defined energy plateaux,
again just as for the $k=2A$ flux tubes. 

As for the $k=2A$ case, it is worth quantifying how well we  can constrain 
the L\"uscher correction with the $k=3A$ ground state. Here we find 
\begin{equation}
aE_{3A}(l) =\sigma_{3A} l -c_{eff} \frac{\pi}{6l} \quad ; \quad c_{eff} =
1.11(11) \ \mathrm{for} \ l \surd\sigma_{3A} \geq 2.3,
\label{eqn_corML3A}
\end{equation}
which is again a usefully accurate test of this universal coefficient.

We turn now to the lightest excited states in the $p=0$ sector, as displayed 
in Fig.~\ref{fig_EQ0k3A_n6f}. Comparing to  Fig.~\ref{fig_Eq0k2a_n6f}
we see that the phase space for the first excited flux tube to decay is smaller 
here and indeed at larger $l$ it is stable. This is perhaps why its energy,
particularly  at small $l$, is closer to Nambu-Goto
than in the $k=2A$ case. And also why the effective energies displayed in
Fig.~\ref{fig_Eeffk3Aq0ex_n6f} show clear plateaux even for $l/a=16$, in
contrast to the $k=2A$ case in Fig.~\ref{fig_Eeffk2Aq0ex_n6f}. 
(Note that for $l/a=16$ we use
our full $k=3$ basis, which means that state includes a very slight
admixture of $k=3M$ and $k=3S$.) The decay
thresholds are indicated on the right hand axis of 
Fig.~\ref{fig_Eeffk3Aq0ex_n6f}, and it is clear that the low-$l$ plateaux 
take very different values. This provides us with a quite clean example
of an excitation of a bound state flux tube where we can ignore
the (slight) instability of the state. It is therefore interesting
to compare this to the corresponding excitation of the fundamental flux
tube in Fig.19 of
\cite{AABBMT_fd3}.
We see that the deviation from Nambu-Goto is indeed very much larger here,
and this must be due to the  bound state structure of this flux tube.
We note that a similar analysis applied to the first $P=-$ excitation 
with  $P=2\pi/l$ leads to very similar conclusions.

We turn now to the heavier $k=3M$ states. We plot in Fig.~\ref{fig_EQgsk3M_n6f}
the ground states with the lowest longitudinal momenta. Once again these
particular states agree very well with the Nambu-Goto predictions. However
we see that they are now slightly above the decay threshold and so  
will be unstable but apparently not enough to affect the extraction of, 
for example, the absolute ground state as we see in Fig.~\ref{fig_Eeffk3Mq0_n6f}.
(Again we use the full $k=3$ basis for the $l/a=16$ ground state.)
However the $p=0$ excited states are very unstable and we are unable to identify
useful plateaux.

The $k=3S$ states are much heavier and we can only estimate energies for
the ground state $p=0, 2\pi/l$ states, as shown in Fig.~\ref{fig_EQgsk3S_n6f}.
Again we see rough agreement with Nambu-Goto, but now
the decay phase space is large -- becoming very large for 
large $l$. We show the effective energies for the absolute ground state in
Fig.~\ref{fig_Eeffk3Sq0n_n6f}. While the plateaux at lower $l$ are
quite clear and are far from the decay thresholds (indicated on the
right hand axis), this is not the case at the largest values of $l$.
(Indeed we do not even attempt to extract an energy for $l/a=48$.)
In the latter cases, while the motivation for our energy estimates should be apparent,
it is not necessarily convincing. Nonetheless the usual agreement with
Nambu-Goto for such states at smaller $l$ is remarkable.

\subsection{adjoint}
\label{subsection_adjoint}

The adjoint flux tube appears in $f\otimes {\bar{f}}$ and should couple to
operators $\mathrm{Tr}_{adj}l_p = |\mathrm{Tr}_f l_p|^2 - 1$, if indeed it exists. There is some 
evidence from the calculation of adjoint potentials that it does indeed exist
and that the adjoint string tension satisfies approximate Casimir scaling
(see e.g.
\cite{CSsu3}
and references therein). Such a flux tube can be screened down to the vacuum
by gluons, but this is suppressed by $1/N^2$, and is in fact negligible except 
for finite volume effects. The latter can either arise if $l$ is small, i.e. 
$l \sim l_c$, or if we consider blocked/smear $l_p$ operators that extend
around the transverse torus. In practice we always include such highly smeared
operators in our calculations, since they (slightly) improve the overlap onto
the ground state of the adjoint flux tube, and we therefore explicitly subtract 
vacuum expectation values in our correlators.

An adjoint flux tube whose string tension satisfies approximate Casimir scaling
will in general be heavier than a pair of fundamental anti-fundamental flux
tubes and can therefore decay into these. (Here there is no large-$N$ suppression.)
Just as with unstable $k$-strings, the important question is whether the adjoint 
flux tube is nearly stable, so that conventional methods for extracting the energy
can be used, or not. We shall be careful to establish whether this is so or not.

In Fig.~\ref{fig_EgsQallAdj_n6f} we plot the energies of the lightest adjoint flux 
tubes with longitudinal momenta $p=0, 2\pi/l, 4\pi/l$. As usual the $p=0$ 
Nambu-Goto fit fixes the string tension $a^2\sigma_{adj}$, and then the Nambu-Goto
predictions for $p\neq 0$ are parameter-free. We observe that, again as usual,
these predictions are remarkably well satisfied all the way down to $l\sim l_c$.
The decay thresholds are indicated and we see that the decay phase space is small,
raising the hope that the decay widths will be negligibly small. Of course 
the statistical errors are quite large here so it is worth extracting the 
`excitation energy' as defined in eqn(\ref{eqn_exq}) to see how well that
is being determined. As we see from Fig.~\ref{fig_DE2NGgsq012adj_n6f} 
the modes carrying momentum are indeed unambiguously the wave-like modes
of a thin relativistic string.

In Fig.~\ref{fig_Eeffadjb_n6f}
we plot the effective energies for the absolute 
ground state. (Energies shifted for clarity.) Horizontal red solid lines indicate
our plateaux estimates, including errors. For small and medium $l$ these are
well determined, but for the largest values of $l$ the states are very massive
and we quickly lose the signal as we go to larger $n_t$. Hence the generous
error estimates in these cases. For comparison we plot the 
$f{\bar{f}}$ threshold energies as horizontal dashed lines (also as points 
on the right hand axis). These are quite close to the plateaux, especially
at small $l$. So we blow up the scale for the latter states in
Fig.~\ref{fig_Eeffadjlb_n6f}. A characteristic feature of effective
energies is that once the error gets large, the estimate of that error becomes
unreliable. This applies to the large $n_t$ decrease or increase in $aE_{eff}(n_t)$
that we see in Fig.~\ref{fig_Eeffadjlb_n6f}. Since our correlators are diagonal,
an increase would violate positivity, and so must be statistical. There is therefore
no reason to take the decreases any more seriously. Given these remarks,
we can see that the $l/a=16$ plateau estimate is consistent with the decay threshold,
while for $l/a\geq 20$ (and unambiguously for $l/a>20$) the plateaux is well
above the threshold. We conclude that the adjoint flux tube does indeed exist as
a distinct and nearly stable `bound state'.  

On the other hand we cannot identify well-defined excited states with $p=0$.
These would have a very large phase space for decay into  $f{\bar{f}}$ flux tubes,
so this is not unexpected. They are presumably analogous to broad resonances,
and will be equally difficult to identify.

\subsection{\underline{84}   $ \ $ and \underline{120}}
\label{subsection_84and120}

In the $f\otimes f\otimes {\bar{f}}$ sector of SU(6), the irreducible
representations with the smallest Casimirs and, we can assume, the
smallest string tensions, are the $\underline{84}$
and  $\underline{120}$. (See the Appendix.) Here  we shall study 
flux tubes carrying flux in these two representations.

Such flux tubes can mix with single fundamental flux tubes, but
this is large-$N$ suppressed and given our experience with the adjoint
flux tube, we shall (usually) ignore this possibility. However the
decay/mixing with 3 (anti)fundamental flux tubes is not large-$N$ suppressed.
And neither is that with a $k=2A$ and an antifundamental, which is even lighter.
In Fig.~\ref{fig_EgsQallR84_n6f} we plot the energies of the
ground state $\underline{84}$ flux tubes with longitudinal momenta
$p=0,2\pi/l,4\pi/l$. The Nambu-Goto predictions are shown as solid red 
curves, with the decay $3f$ and $2A+f$ thresholds indicated by the black 
curves. As usual we extract the string tension from the $p=0$ fit so that 
the $p\neq 0$ predictions are parameter free. We observe that the
agreement is, once again, remarkably good for $p=2\pi/l$ and quite
good for $p=4\pi/l$, where however the states are very massive and it becomes
difficult to identify plausible plateaux. The string tension is comparable
to that for $k=3M$ (see Table~\ref{table_sigma_R}) as is the phase
space for decays. So it is no surprise that, just as for $3M$, we are
unable to obtain energy estimates for $p=0$ excited states. The
ground state however has reasonably clear energy plateaux, as we
see in Fig.~\ref{fig_Eeffr84_n6f}, at least for $20 \leq l/a \leq 40$.
For $l /a\geq 44$ the effective energies are large and disappear rapidly
into the statistical noise as $n_t$ increases, making plateau
identification increasingly subjective. For $l/a=16$ we see no plateau,
and here we see that $E_{eff}(n_t)$ decreases well below the decay
thresholds shown and appears to be asymptoting to a large-$N$
suppressed single $f$ admixture. That this should only occur for our
shortest flux tube, $l/a=16$, is consistent with our earlier
observations about the finite volume effects displayed in
Table~\ref{table_2A2Sover}.

In Fig.~\ref{fig_EgsQallr120_n6f} we plot the ground state energies of 
flux tubes in the $\underline{120}$ representation for $p=0,2\pi/l$.
The  $\underline{120}$ string tension, which we obtain by fitting the $p=0$
values, is almost as large as the $k=3S$ one, and so it is no surprise that
just as in that case we have no useful results for  $p=4\pi/l$ or
for any excited states. Indeed even the $p=0$ effective energy plateaux
are difficult and ambiguous to identify in this case.

Finally we remark that we have also performed some matching
calculations in SU(3) at $\beta=40.0$, which corresponds to about
the same lattice spacing. The corresponding  $f\otimes f\otimes {\bar{f}}$
irreducible representations are the $\underline{6}$ and  $\underline{15}$.
In both cases the energy plateaux are more ambiguous, particularly
where we compare the $\underline{15}$ with the $\underline{120}$ of SU(6).
This may be due to the fact that certain mixings and decays are
less suppressed for SU(3) than for SU(6).

\section{Excited states: massive or stringy?}
\label{section_massiveorstringy}

One of our motivations for studying flux tubes in higher representations
is that we expect such bound states of fundamental flux tubes to have
a low-lying excitation spectrum that contains clear signatures of the binding
scale. This should provide an interesting contrast to the low-lying spectrum of 
fundamental flux tubes which, unexpectedly, shows no sign of the excitation
of the massive modes that one would expect to be associated with an
`intrinsic width' for the flux tube. While one might question the
existence of such an intrinsic width, the existence of a non-zero binding in
the case of, say, the $2A$ flux tube is unambiguous.
This would, most simply, reveal itself in extra excited states,
representing massive rather than the usual stringy massless modes.
Our cleanest spectra in this paper are for $k=2A$ and $k=3A$ so we shall focus
on these. So does the $k=2A$ $p=0$ spectrum shown in 
Fig.~\ref{fig_Eq0k2a_n6f} reveal any massive modes that are additional
to the stringy excitations which, at large $l$, tend to the Nambu-Goto 
curves? (The same observations apply to the $k=3A$ spectrum.)
Since the low-lying excitation spectrum of fundamental flux tubes
appears to contain only stringy states and no massive modes, it is
interesting to compare our $k=2A$ spectrum to the fundamental one 
shown in Fig.12 of
\cite{AABBMT_fd3}.
The immediate question this comparison raises, as pointed out in our 
earlier study of $k=2$ flux tubes in
\cite{AABBMT_k2d3},
is whether the first excited $k=2A$ state might be a massive mode, with the 
second excited state being the first excited stringy mode and the next two
$P=+$ excited states eventually tending to the second Nambu-Goto level? 
(We have not shown higher $P=+$ excited states in Fig.~\ref{fig_Eq0k2a_n6f},
but they are there.) Or it might be that the large
deviations from Nambu-Goto are largely driven by the `unstable'
character of these flux tubes, and that otherwise the first excited state
is much like the fundamental one. However this possibility appears to be
contradicted by our results in this paper for the much more stable $k=3A$ 
states, plotted in  Fig.~\ref{fig_EQ0k3A_n6f}, which show similarly
large deviations from the Nambu-Goto predictions. Or again, it might be 
that we are seeing here the mixing of modes, enhanced by the 
existence of intermediate states that are not far from threshold.
This could be the mixing of nearby stringy modes, or of a stringy
mode with a massive mode  - which would also imply the presence
of an extra mode.
 
So we want to ask if the first excited states in the $k=2A$ and $k=1$
cases are the `same' or not. It is of course not possible to answer
this question unambiguously, and we choose to address it in the same way
as we did in 
\cite{AABBMT_k2d3}.
The idea is that if this state is indeed an approximate Nambu-Goto-like 
string excitation then we would expect its wave-functional
to have the appropriate `shape'. What that `shape' should be, in
terms of our highly blocked/smeared link matrices, is not at all 
evident, but it is something we do not need to know because we can simply 
compare it to the wavefunctional of the first excited $k=1$ state,
which we have good reason to think of as being stringy..

The way we make this comparison is as follows. Let 
$\{\phi_i; i=1,...,n_o\}$ be our set of winding flux tube operators,
with  $P=+$ and $p=0$. These operators are group elements, 
not yet traced, and may be in any representation of SU($N$).
Suppose the flux is in the representation $R$. When we
perform our variational calculation over this basis, we obtain
a set of wavefunctionals, $\Phi^n_{R}$, which are an approximation 
to the corresponding eigenfunctions of the Hamiltonian. 
Unfortunately we cannot simply compare $R=f$ and $R=2A$ states by 
calculating their overlap: it will vanish because of the center 
symmetry. So instead we proceed as follows
\cite{AABBMT_k2d3}. 
We write the wavefunctionals as linear combinations of 
our basis operators:
\begin{equation}
\Phi^{n}_{R}
=
\sum^{n_o}_{i} b^{n}_{R,i} c_{R,i}{\rm Tr}_{R} (\phi_i)
\equiv
\sum^{n_o}_{i} b^{n}_{R,i} {\rm Tr}^{\prime}_{R} (\phi_i)
\label{phie2}
\end{equation}
choosing the coefficients $c_{R,i}$ to satisfy the normalisation condition
\begin{equation}
\langle {\rm Tr}^{\prime\dagger}_{R} (\phi_i(0))
{\rm Tr}^{\prime}_{R} (\phi_i(0))\rangle
= 1
\label{phinorm}
\end{equation}
so as to ensure that a comparison of the coefficients $ b^{n}_{R,i}$ 
between different representations ${R}$ can be meaningful.
The idea is that the coefficients $b^{n}_{R,i}$
encode the `shape' of the state corresponding to the wavefunctional,
because they multiply the same operators, albeit in different
representations, and with a common normalisation. So making the
simple substitution 
\begin{equation}
{{\Phi}}^n_{2A}
=
\sum^{n_o}_{i} b^{n}_{{2A},i} {\rm Tr}^{\prime}_{2A} (\phi_i)
\quad \longrightarrow \quad
{\tilde{\Phi}}^n_{2A}
=
\sum^{n_o}_{i} b^{n}_{{2A},i} {\rm Tr}^{\prime}_{f} (\phi_i)
\label{phie3}
\end{equation}
we can compare our excited $k=1$ and $k=2A$ wavefunctionals by
comparing ${\tilde{\Phi}}^n_{2A}$ with the fundamental wavefunctionals, 
${{\Phi}}^n_{f}$. This we can do by calculating the overlap 
\begin{equation}
O_{n^\prime,n}=
\frac{ \langle {\Phi^{n^\prime\dagger}}_f {\tilde{\Phi}^n}_{2A} \rangle}
{\langle{\Phi^{n^\prime\dagger}}_f \Phi^{n^\prime}_f \rangle^{1/2} 
\langle {{\tilde{\Phi}}}^{n\dagger}_{2A} \tilde{\Phi}^{n}_{2A} \rangle^{1/2} }
\label{overlap}
\end{equation}
(all operators at $t=0$)
which we assume provides us with a measure of the similarity between 
the original state ${{\Phi}}^n_{2A}$ and  the state ${{\Phi}}^{n^\prime}_{f}$.

Even if one accepts this method of comparison, there are some 
important caveats.
The variational calculation is performed over a limited basis,
so the ${{\Phi}}^n_{f}$ are only approximate energy eigenfunctionals.
And the level of approximation will generally be different for
different representations (and states). Thus the comparison is
inevitably approximate. Again, we note that the operator basis   
$\phi_i$ varies with $l$ (in lattice units). So we perform the 
comparison of $f$ and $2A$ states at the same $l$. However ideally we 
should also compare at the same string tension i.e. at different
lattice spacings such that $a\surd\sigma_{2A}=a^\prime\surd\sigma_{f}$
and hence different $\beta$. Because of the additional costs
we have not done so here, and this also makes the comparison
approximate.

Given the approximate and heuristic nature of this method, we need 
to test it in a case where we are confident that we know the answer. 
This is the case for the absolute ground state. So in 
Fig.~\ref{fig_over2AAdjl32_gs} we display the above overlaps 
of the variational ground states of the $k=2A$ and adjoint flux tubes 
onto the lightest 20 fundamental variational eigenfunctionals, all on $l/a=32$
lattices. In Fig.~\ref{fig_over2AAdjl64_gs} we do the same 
on a $l/a=64$ lattice. The result is clear-cut, both for
stable and unstable flux tubes, and for both lengths: we observe that 
the method works very well in producing an almost exclusive 
overlap onto the $f$ ground state. This is in fact
representative of all our results for the absolute ground state,
even where the state is unstable, and this gives us some confidence in 
this method.

We turn now to the $k=2A$ $p=0$ first excited state. In
Fig.~\ref{fig_over2Al3264_ex1} we show the overlap of ${\tilde{\Phi}}^{n=1}_{2A}$
onto the lowest ${{\Phi}}^{n}_{f}$, for $l/a=32$ and $l/a=64$ lattices. 
In Fig.~\ref{fig_over3Al3264_ex1} we do the same
for the corresponding $k=3A$ state. While the largest overlap is indeed
on the first excited fundamental state, there is also a small but visible 
overlap onto the 2nd excited stringy $f$ state, ${{\Phi}}^{n=2}_{f}$,
and this is very similar for the $k=2A$ and $k=3A$ states and for both
flux tube lengths. While the comparison is not as unambiguous as for 
the ground state, it is hard to avoid the conclusion that this state
is definitely not some new massive mode excitation. Rather it appears to
be largely the first excited stringy mode, with a modest admixture of 
the second. The shift in energy away from Nambu-Goto might be largely the 
result of this mixing. We also note that the mixing appears to become smaller 
as $l$ increases from $l/a=32$ to $l/a=64$ and the energy approaches that
of the Nambu-Goto prediction. We remark that all this confirms that the
first excited state at lower $l$ does indeed asymptote to the first excited Nambu-Goto
energy level, and does not cross the latter somewhere between $l/a=52$
and $l/a=64$ -- an alternative possibility that we discussed earlier,  in
Section~\ref{subsection_k2}, when considering Fig.~\ref{fig_Eq0k2a_n6f}.

It is interesting to contrast this with what one finds for the 2nd and
3rd excited states in the $k=2A,3A$ and $p=0,P=+$ sectors. Typical examples 
are shown in Figs.\ref{fig_over2Al32_ex2ex3} and \ref{fig_over2Al52_ex2ex3}.
Here it is hard to draw any conclusion. While the dominant overlap
is onto the corresponding fundamental excited state, there is a large
projection on other states as well. It certainly appears possible that
some new massive mode either dominates or is mixed into one or both of
these states.

\section{Discussion and conclusions}
\label{section_conclusion}

In this paper we have calculated the low-lying spectrum of closed flux 
tubes in various representations, with the length of the flux tube stabilised 
by closing it around a spatial torus. We had several motivations for 
this study. 

One is to compare the resulting spectrum to simple effective 
string actions, just as we did in our earlier work on fundamental flux tubes
\cite{AABBMT_fd3}. 
Since higher representation flux tubes can be thought of as bound states of 
(anti)fundamental flux tubes, the massive excitation modes associated with
that binding should leave a signature in the spectrum. In the case of
fundamental flux tubes we found no trace at all of massive non-stringy modes
and our hope was to find something different here. 

Of course only a few of
these flux tubes are stable against decay and are real bound states. 
Recall that only some decays are large-$N$ suppressed by large-$N$ 
counting arguments. For example the
decay of an adjoint flux tube to the vacuum (plus glueballs) 
is suppressed at $N=\infty$, but 
its decay into a pair of noninteracting fundamental and anti-fundamental
flux tubes need not be, since 
$\mathrm{Tr_{adj}}l_p = \mathrm{Tr_f}l_p\mathrm{Tr_f}l_p^\dagger -1$. 
(Although the dynamics may of course suppress such decays.) Most
higher representation `flux tubes' are unstable at large $N$, in this
sense, and it is not a priori clear if they exist in the same way as unstable 
`resonant' particle states exist. While there is evidence for the 
existence of such flux tubes when attached to appropriate sources
\cite{CSsu3}, 
their stability in that case is usually ensured (for the relevant length scales)
by the fact that screening the sources by gluons costs extra energy. Our closed 
flux tubes are not protected from being screened, and decaying, in this way.
So one of the things we wished to learn is which of the unstable flux tubes were 
stable enough that one could analyse them by conventional methods, and which
needed new methods and what those new methods might be. 

A closely related question is whether it makes sense to classify flux tubes 
according to the irreducible representations of SU($N$), given that the 
vacuum contains adjoint gluons that can screen and mix the representations 
of sources and flux tubes. Earlier calculations have provided
evidence that this is indeed the case for $k$-strings
\cite{BBMT_kd3,AABBMT_k2d3}, 
and in Table~\ref{table_2A2Sover} we have provided similar evidence for all
the representations being considered here. Apart from finite volume effects
(both in the transverse size and in $l$) the screening appears to be (almost?) 
exact. Why this should be so, at what is not a very large value of $N$, 
is an interesting puzzle that is being investigated by us, more 
systematically, elsewhere
\cite{AAMT_overlap}.

Our first conclusion, discussed in Section~\ref{section_results},
is that the absolute ground state and the lightest states with non-zero 
longitudinal momenta are accurately described by the free 
string expression in eqn(\ref{eqn_EnNG}) all the way down to our lowest
values of $l$, which are very close to the minimal possible 
flux tube length at $l = 1/T_c$. This is clearest for the very stable
flux tubes in the $k=2A$ and $k=3A$ representations, which are our
most accurate calculations (see Figs~\ref{fig_EgsQallk2a_n6f} 
and \ref{fig_EQgsk3A_n6f}) 
but it is also the case, within larger errors, for all representations, 
including flux tubes that could be very unstable. (See
 Figs.\ref{fig_EgsQallk2s_n6f},~\ref{fig_EQgsk3M_n6f},~\ref{fig_EQgsk3S_n6f},~\ref{fig_EgsQallAdj_n6f},~\ref{fig_EgsQallR84_n6f},~\ref{fig_EgsQallr120_n6f} for
the  2S, 3M, 3S, adjoint, 84, and 120 representations respectively.) This
is of course just what has been observed for
the corresponding fundamental flux tube states in
\cite{AABBMT_fd3}. 

In the case of the $p=0$ ground state the stringy corrections are
small all the way down to $l\surd\sigma \sim 1$ (basically because
they arise  from the zero-point energies) and so an expansion of the
energy in powers of $1/l\surd\sigma$ is convergent over our whole
range of $l$. So it is fair to claim
that the close agreement we are seeing with Nambu-Goto down to
small values of $l\surd\sigma$ is a prediction of the known universal 
corrections to the linear $\sigma l$ piece
\cite{OAZK,OA,JM,LW}
-- at least when the flux tube is stable. In fact,  as discussed in
Section~\ref{subsection_k2}, what we are able to confirm, within our 
statistical errors, is the presence of 
the linear piece and, in some cases, the $-\pi/6l$ universal
L\"uscher correction, but not really any more than that. (See e.g. 
Fig.~\ref{fig_DENGgsq0P+k2a_n6f}.) Our most accurate
spectra, for the $k=2A$ and $k=3A$ representations, allowed us to confirm
the universal value of this coefficient at the $\pm 10\%$ level, which
is a usefully accurate result for these bound-state flux tubes. 

The $p\neq 0$ ground states are another matter. Here we can confirm, 
in some cases very accurately, the excitation energy $=|p|$ of the 
massless excitation that carries the momentum on the background 
flux tube, as shown in 
Figs.\ref{fig_DE2NGgsq0012k2a_n6f},~\ref{fig_DE2NGgsq0012k2s_n6f},~\ref{fig_DE2NGgsq012adj_n6f} 
for the $2A$, $2S$ and adjoint flux tubes respectively. 
In fact  our calculations are accurate enough to confirm the presence 
of the additional zero-point energy, as discussed in Section~\ref{subsection_k2}
and displayed in Fig.\ref{fig_DE2NGgsq0012k2a_n6f}.
Indeed we saw that in any non-stringy attempt to describe these spectra, 
the  particle excitation carrying the non-zero momentum will have a
mass that is constrained by our calculated spectrum to be very much 
smaller than the known mass gap of the bulk space-time theory
\cite{BLMT_d3,MT_98}. 
Thus such a (presumably massless) excitation must exist on the flux 
tube rather than in the bulk, and will thus arise from an effective 
string action.

That we observe a (near) free-string behaviour,  even when the flux 
tube is very short, is more surprising for the  $p\neq 0$ ground states
than  for the $p=0$ ground state.
This is because the expansion in powers of  $1/l\surd\sigma$ diverges
 at quite large $l$ for $p\neq 0$, so we cannot use
universality arguments to predict the  $p\neq 0$ spectrum at small $l$ 
in the way we could for $p=0$. All this parallels what 
has previously been seen for the fundamental flux tube 
\cite{AABBMT_fd3}.
There we  suggested
\cite{AABBMT_fd3}
that what these ground states have in common is that they all have a phonon content 
that is either all left moving or all right moving, so that all phonon 
subenergies are at threshold and it is plausible that the interaction of 
these Goldstone bosons will vanish there, removing at least one
possible source of corrections to the free-string result. We note that
a very recent and much more complete analysis of phonon scattering in a finite
volume comes to a similar conclusion
\cite{SDRFVG_13}.
Our results here, with higher representation flux tubes, are consistent with this picture.  
We also remark that, 
as shown in Fig.\ref{fig_EgsQallk2a_n6f},  a `minimalist' model where 
the ground state with $p\neq 0$ consists of a $p=0$ flux tube together 
with massless noninteracting particles sharing $p$, gives predictions
very close to Nambu-Goto. (In contrast to a large discrepancy for other 
states that involve both right and left movers.) It may thus be that, 
to a first approximation, this part of the spectrum is independent of the
model used (within limits). But this is only a speculation.

By contrast other states, for example the $p=0$ excited $k=2A$ and $k=3A$ 
states shown in Figs~\ref{fig_Eq0k2a_n6f} and \ref{fig_EQ0k3A_n6f}, 
show very large deviations from Nambu-Goto, making it hard to say to what 
extent that model provides any kind of first approximation
to this part of the spectrum. We recall that in the case of fundamental 
flux tubes the corresponding corrections are significant but small, 
with a rapid approach to the free-string spectrum as $l\uparrow$
\cite{AABBMT_fd3}. 
There are two obvious differences between the fundamental flux tube
and the ones here. Firstly, here we have a binding dynamics which
may perturb the spectrum through its excitations. Secondly most of the
states are not only unstable at finite $N$ but remain so in the $N\to\infty$ 
limit. If mixing is the dominant effect, then the states will contain
non-stringy components, while if instability is the dominant effect
then they will be stringy but resonance-like.  In this context we note
that at the larger values of $l$ in  Fig.\ref{fig_EQ0k3A_n6f}, the 
lightest $p=0$ excited states become stable, in contrast to the
case of $k=2A$ shown in  Fig.\ref{fig_Eq0k2a_n6f}. Nonetheless
the deviations from Nambu-Goto are not significantly different 
in the two cases. This suggests that instability is not the main
reason for the large deviations we see for the second excited state
at these values of $l$.

The striking difference between the simple stringy behaviour of
the $p\neq 0$ ground states, even when these are unstable,
and the messy behaviour of the unstable excited states may
be due to the fact that in the former case, unlike the latter, the 
phonons will not catalyse the decay of the flux tube because they
have zero subenergies
\cite{decays}.
Although this may not be the only source of flux tube decay,
it may be a significant factor.

To make some progress
in  identifying the nature of the $p=0$ excited states we introduced in
eqns(\ref{phie2}-\ref{overlap}) a heuristic measure
\cite{AABBMT_k2d3} 
for comparing states in different 
representations. Applied to the $p=0$ ground states it confirmed unequivocally
that these are just the same as the unexcited flux tube in the fundamental 
ground state, and that this is so for all the representations we consider here. 
See for example Figs~\ref{fig_over2AAdjl32_gs} and  
\ref{fig_over2AAdjl64_gs} for the $k=2A$ and adjoint 
cases. This is as expected, and motivates the use of the measure for the
more controversial excited states. For the first excited $k=2A$ $p=0$ state, 
as shown in Fig.~\ref{fig_over2Al3264_ex1}, the result is less unequivocal 
but points to it being
quite similar to the corresponding fundamental state and becoming more so as 
$l\uparrow$. We note that what we show in the figures is the overlap-squared, 
whereas it is possible that the energy shift contains pieces proportional to the 
overlap, which is larger and could produce a significant shift in the energy.
For the second and third excited $p=0$ states, analysed in 
Figs~\ref{fig_over2Al32_ex2ex3} and  \ref{fig_over2Al52_ex2ex3}, 
we see states that are very unlike the corresponding
fundamental ones. This suggests that while the ground and first excited $p=0$
states in Fig.~\ref{fig_Eq0k2a_n6f} are indeed (mostly) stringy, the higher 
states may well include a large admixture with massive modes. 

Because nearly all our states are unstable, it is important to approach our
energy estimates as critically as possible, and we have attempted to do so.
(At the risk of being tedious.)
One usually obtains an energy from a correlation function by
calculating the effective energy, $aE_{eff}(t) = -\ln C(t+a)/C(t)$,
and identifying a `plateau' for $t\geq t_0$ where $t_0$ has to be small
if the result is to be usefully accurate. The stable $k=2A, 3A$ $p=0$ ground 
states provide our benchmark for an unambiguous calculation, as shown in
Figs~\ref{fig_Eeffk2Aq0_n6f},~\ref{fig_Eeffk3Aq0_n6f}. 
As $l\uparrow$ the energy increases $\propto l + O(1/l)$
and the signal, $\sim \exp\{-aE(l)n_t\}$, drops into the statistical noise 
at smaller $t=an_t$, so decreasing the useful extent of the plateau. Apart
from this there is no ambiguity in extracting energies. The $p\neq 0$
ground states are similar, although the energies are larger so, once
again, reducing the useful energy plateau. States that are unstable but
with a small decay width should, by continuity, produce a plateau
at intermediate $t$, which eventually sinks to the decay threshold.
(As long as we remain within the same limited class of operators that
are designed to project onto single flux tubes.) If the decay width
is large, we will lose all sign of a plateau. Examples of the former
are the adjoint ground state in Figs~\ref{fig_Eeffadjb_n6f} 
and \ref{fig_Eeffadjlb_n6f} and, less stable, the $k=2S$ ground 
state in Fig.~\ref{fig_Eeffk2Sq0_n6f}. A totally unstable state is
the first excited $k=2S$ $p=0$ state shown in 
Fig.~\ref{fig_Eeffk2Sq0ex_n6f}. For obvious reasons
we have not attempted to extract an energy for this state. This is to be
contrasted with the first excited $k=2A$ $p=0$ state which we show 
in Fig.~\ref{fig_Eeffk2Aq0ex_n6f}. While there is no plateau for $l/a=16$, 
one can just about attempt to discern one for $l/a=20$ 
and more easily for higher $l$. We attempted heuristic 
unstable particle fits to such effective masses, using decay channels weighted 
with a Breit-Wigner for the unstable state as in eqn(\ref{eqn_corBW}). 
Examples are in Fig.~\ref{fig_Eeffk2Aq0ex_l16_n6f}
and  Fig.~\ref{fig_Eeffk2Sq0ex_l16_n6f}. The interesting feature of 
such fits is that the 
true energy is not to be found by looking at $E_{eff}(t)$ at large $t$, but
is typically close to the value of   $\lim_{t\to 0}E_{eff}(t)$ i.e. it is very
much larger. This could significantly reduce the large discrepancy between
Nambu-Goto and our observed $p=0$ excited states. This highlights the dangers
in using conventional methods to make energy estimates when the state
is not stable or very nearly so.

Despite our uncertainty concerning the existence of many of the excited states,
the existence of the absolute ground states, whether stable or unstable,
appears to be quite unambiguous. And this is often also true of the
ground states with $p\neq 0$. Moreover even when the flux is in our rather
exotic representations, these particular states have the simple stringy excitations 
of a free string theory, even as $l$ decreases close to its physically 
minimum value. The spectrum of other excited states is more complex and
there appears to be plenty of room there for the massive modes that might 
arise from the binding energy in these states.

\section*{Acknowledgements}
During the course of this work, MT participated in the {\it{Recent 
Advances in Numerical Methods for Field Theory and Gravity}}
Workshop at the KITP, Santa Barbara,
and in the {\it{New Frontiers in Lattice Gauge Theory}}  Workshop 
at the Galileo Galilei Institute, Florence, and acknowledges
useful discussions with a number of participants. Part of this work 
was also done during the {\it{Large-N gauge theories}} Workshop at the 
Galileo Galilei Institute, Florence, and AA and MT would like to thank the 
organisers for the warm hospitality and many of the participants for useful 
discussions. This paper is a continuation of a project whose earlier stages were
in collaboration with Barak Bringoltz whom we thank, with pleasure,
for his role in helping to develop many of the 
ideas and techniques used in this paper.
The computations were carried out on computers in Oxford Theoretical 
Physics funded by EPSRC and Oxford University.

\clearpage


\appendix

\section{Representations and Casimirs}
\label{section_appendix_reps}

In this Appendix we describe in detail how to calculate flux tubes
in the representations of interest to us in this paper, and how to
determine some of their group theoretic properties. 

Consider a Polyakov loop 
\begin{equation}
l_p=\prod_{n_x=1}^{n_x=L_x} U_x(n_x)
\label{eqn_poly_b}
\end{equation}
that winds once around the (spatial) $x$-torus of length $L_x$
in lattice units. (We suppress other co-ordinates.) 
If we take the trace in the representation $R$, 
then the   operator $\mathrm{Tr}_R\{l_p\}$ is a candidate operator for projecting 
onto winding flux tubes carrying flux in the representation $R$. 
One can deform the Polyakov loop in eqn(\ref{eqn_poly_b}) so as
to obtain winding operators that can be used to form flux tubes
with non-trivial quantum numbers. These operators can be used as a 
basis for a variational calculation of the spectrum as described in
the text. We use here the same basis as we used for the fundamental
flux tube in
\cite{AABBMT_fd3},
and we refer to that paper for a list of the operators. 

The open version of such a flux tube would connect sources in
the representations $R$ and $\overline{R}$. But because the vacuum contains 
gluons a flux tube in representation $R$ may evolve into one in
$R^\prime$ by gluon screening  if $R$ and $R^\prime$ can
be connected by a product of adjoints. And of course a flux tube in
$R$ may evolve into a product of flux tubes carrying fluxes $R_1$, $R_2$, ... if
$R$ appears in the product $R_1\otimes R_2\otimes ...$. 
Equally an excited flux tube may decay into a lower lying flux tube 
of the same representation plus colour singlet glueballs. The same
remarks apply to a closed flux tube winding around the spatial torus.
In most cases such a flux tube will not be absolutely stable and 
whether it exists at all, in the sense that an unstable but narrow resonance
exists, will depend on the dynamics.    

Note that some decays will be suppressed by the usual
`kinematic' large-$N$ counting arguments, for example
the decay of a sufficiently excited flux tube state into its ground 
state together with a glueball. Other decays are not suppressed in
this way, e.g. the decay of an adjoint flux tube into a pair of 
fundamental and anti-fundamental flux tubes. (Although it may be that 
the detailed dynamics suppresses such a decay.) The latter is
analogous to the expected `falling apart' as $N\to\infty$ of a molecular 
hadronic state (such as, perhaps, the inverted nonet scalars at 
$\leq 1\mathrm{GeV}$), while the former is analogous to the $\rho$ meson
becoming stable in that limit.  

Consider the fundamental representation $f$. Any representation $R$ will 
appear in a product of a number of $f$ and ${\overline{f}}$. Suppose that 
the number of these is $N_f$ and $N_{\overline{f}}$ respectively, then we define
the ${\cal{N}}$-ality of the representation to be
\begin{equation}
k =  N_f - N_{\overline{f}}.
\label{eqn_nality}
\end{equation}
and we use the generic name of a `$k$-string' for such a flux tube.
The reason for focusing on the ${\cal{N}}$-ality 
is that for an SU($N$) gauge theory $k$ is conserved 
(module $N$) under gluon screening (since gluons are adjoints
carrying $k=0$). So for $N\geq 4$ the lightest
$k=2$ flux tube is an absolutely stable state, as is the lightest
$k=3$ flux tube for $N\geq 6$. Of course it may be that such a $k=2$ `flux 
tube' consists of nothing more than two $k=1$ flux tubes. Whether it 
does or not is a dynamical question which, in fact, has been answered 
(see e.g.
\cite{BLMT_strings,BBMT_kd3}),
and we know that the lightest $k=2$ and $k=3$ strings are flux tubes 
in their own right, which are strongly bound, stable states in SU(6)
\cite{BBMT_kd3}.

The irreducible representations we consider in this paper include
the totally anti-symmetric and symmetric  of $f\otimes f$
(referred to as $k=2A$ and $k=2S$), and the  anti-symmetric, mixed 
and symmetric of $f\otimes f\otimes f$ (referred to as $k=3A$, $k=3M$ 
and $k=3S$). The construction and properties of these are obtained in 
the Appendices of
\cite{BLMT_strings}
to which we refer the reader for details, in particular for the 
derivation of the appropriate operators, and for the values of the 
quadratic Casimir, 
\begin{equation}
C_2(R) = \mathrm{Tr_R}T^aT^a
\label{eqn_casimirR}
\end{equation}
where $R$ is the representation and  the $T^a$ 
are the generators of the group. 

We also consider in this paper the following representations: the adjoint, 
which is $k=0$ and is the non-singlet piece of $f\otimes{\overline{f}}$, and
the $\underline{84}$ and $\underline{120}$
which are both $k=1$ and which arise in $f\otimes f\otimes {\overline{f}}$.
These are the representations we shall now describe in more detail. 

A standard and efficient method for dealing with the representations of 
SU($N$) is Young tableaux. For the derivation and rules of use we refer to
\cite{IN}.
We will label a tableau by $Y(\lambda_1,\lambda_2, ...)$ where
$\lambda_i$ is the number of boxes in the $i'th$ row of the tableau.
We recall the rule that $\lambda_i\geq \lambda_{i+1}$, and we only 
show the $\lambda_j$ for rows $j$ containing at least one box. So, for
example, $f$ corresponds to $Y(\lambda_1=1)$, $\overline{f}$ corresponds to
$Y(\lambda_1=1,\lambda_2=1, ... ,\lambda_{N-1}=1)$, while the adjoint is 
given by $Y(\lambda_1=2,\lambda_2=1, ... ,\lambda_{N-1}=1)$.
We also recall that the dimension $d_R$ of the representation $R$ 
corresponding to the tableau $Y(\lambda_1,\lambda_2, ...)$ is given by
\cite{IN}
\begin{equation}
d_R 
=
\frac{\prod^N_{i<j}(l_i-l_j)}
{(n-1)!(n-2)!... 1!} 
\quad ; \quad l_k=\lambda_k+N-k \ , \ k=1,...,N
\label{eqn_dimR}
\end{equation}
and that its quadratic Casimir is given by
\cite{BLMT_strings}
\begin{equation}
C_2(R) 
=
\frac{1}{2}
\left(n_bN+\sum_{i=1}^{n_r}\lambda_i(\lambda_i+1-2i)-\frac{n_b^2}{N}\right).
\label{eqn_C2R}
\end{equation}
where $n_b=\lambda_1+\lambda_1+...$ is the total number of boxes in the
tableau and $n_r$ is the number of rows.

Applying eqns(\ref{eqn_dimR},\ref{eqn_C2R}) to the fundamental tableau,
$Y(\lambda_1=1)$, gives
\begin{equation}
d_{f}=N \quad;\quad C_2(f)=\frac{N^2-1}{2N}
\label{eqn_fun}
\end{equation}

The adjoint representation is obtained from $f\otimes{\overline{f}}
=\underline{1}+\underline{adj}$. The singlet corresponds to adding the
single $f$ box to the bottom of the $\overline{f}$ column of boxes,
while the adjoint is obtained by adding it to the first row. Applying 
eqns(\ref{eqn_dimR},\ref{eqn_C2R}) to the resulting adjoint tableau,
$Y(\lambda_1=2,\lambda_2=1, ... ,\lambda_{N-1}=1)$, gives
\begin{equation}
d_{adj}=N^2-1 \quad;\quad C_2(adj)=N
\label{eqn_adj}
\end{equation}
For SU(6) we see that $d_{adj}=35$.

Consider now the product $\overline{f}\otimes f \otimes f$. As we saw 
above the first product gives us 
$\overline{f}\otimes f=\underline{1}+\underline{adj}$. Adding the
second $f$ box to the singlet just gives us an $f$ again, while
adding it to the adjoint $Y(\lambda_1=2,\lambda_2=1, ... ,\lambda_{N-1}=1)$
gives us  $R_a=Y(\lambda_1=3,\lambda_2=1, ... ,\lambda_{N-1}=1)$ if we add it
to the first row and
$R_b=Y(\lambda_1=2,\lambda_2=2,,\lambda_3=1, ... ,\lambda_{N-1}=1)$
if we add it to the second, as well as a second $f$ by adding it to the 
bottom of the first long column (which is then of length $N$ and
so can be dropped).
Applying eqns(\ref{eqn_dimR},\ref{eqn_C2R}) to these tableaux gives,
\begin{equation}
d_{R_a}=\frac{1}{2}N(N+2)(N-1) 
\quad;\quad 
C_2(R_a)=\frac{(3N-1)(N+1)}{2N} = C_2(f)\frac{3N-1}{N-1}
\label{eqn_Ra}
\end{equation}
and
\begin{equation}
d_{R_b}=\frac{1}{2}N(N+1)(N-2) 
\quad;\quad 
C_2(R_b)=\frac{(3N+1)(N-1)}{2N} = C_2(f)\frac{3N+1}{N+1}.
\label{eqn_Rb}
\end{equation}
For SU(6) we can label these representations by their dimensions, i.e.
\begin{equation}
R_a=\underline{120} \ \ ; \ \ R_b=\underline{84} \qquad: \mathrm{for \ SU(6)}
\label{eqn_su6}
\end{equation}
and evaluating the quadratic Casimirs we obtain the entries in 
Table~\ref{table_sigma_R}. 

To calculate the spectrum of a flux tube in representation $R$
we want correlators of operators  $\mathrm{Tr_R}l_p$ where $l_p$
is some loop winding (once) around the appropriate spatial torus,
the simplest example being that in eqn(\ref{eqn_poly_b}). Since
our Monte Carlo generates group elements $U_l$ that are in the 
fundamental representation, we need to express $\mathrm{Tr_R}l_p$
in terms of  $\mathrm{Tr_f}l_p$. This can be done by taking 
products of group elements and imposing the (anti)symmetry
constraints on the indices that are encoded in the corresponding
Young tableaux, as carried out explicitly in
\cite{BLMT_strings}
for the $2A,2S$ and $3A,3M,3S$ representations. Often we can
employ a short cut. For example if we want $R$ and we find that
$R_1\otimes R_2 = R_3 \oplus R$ then we can use the fact that
$\mathrm{Tr_{R_1}}l \times \mathrm{Tr_{R_2}}l = 
\mathrm{Tr_{R_3}}l + \mathrm{Tr_{R}l}$. For example, we know that
the adjoint satisfies 
$f\otimes\overline{f} = adj \oplus \underline{1}$
(where $\underline{1}$ is the singlet) so 
\begin{equation}
\mathrm{Tr_{adj}}l
=
\mathrm{Tr_{f}}l\mathrm{Tr_{\overline{f}}}l - 1
= 
|\mathrm{Tr_{f}}l|^2 -1.
\label{eqn_opadj}
\end{equation}
We also want such an expression for the representations $R_a$ and $R_b$.
One can easily see that $2S\otimes\overline{f}=R_a\oplus f$ giving
\begin{equation}
\mathrm{Tr_{R_a}}l
=
\mathrm{Tr_{2S}}l\mathrm{Tr_{f}}l^\dagger-\mathrm{Tr_{f}}l
=
\frac{1}{2}\left(\mathrm{Tr^2_{f}}l + \mathrm{Tr_{f}}l^2\right)
\mathrm{Tr_{f}}l^\dagger-\mathrm{Tr_{f}}l
\label{eqn_opRa}
\end{equation}
and that  $2A\otimes\overline{f}=R_b\oplus f$ giving
\begin{equation}
\mathrm{Tr_{R_b}}l
=
\mathrm{Tr_{2A}}l\mathrm{Tr_{f}}l^\dagger-\mathrm{Tr_{f}}l
=
\frac{1}{2}\left(\mathrm{Tr^2_{f}}l - \mathrm{Tr_{f}}l^2\right)
\mathrm{Tr_{f}}l^\dagger-\mathrm{Tr_{f}}l
\label{eqn_opRb}
\end{equation}
where we use the fact that $\mathrm{Tr_{\overline{f}}}l=\mathrm{Tr_{f}}l^\dagger$
and we obtain $\mathrm{Tr_{2S,2A}}l$ from
\cite{BLMT_strings}.
The above expressions are for general $N$ and for SU(6) give us
the appropriate operators for the adjoint, $\underline{120}$ and
$\underline{84}$ respectively.


%
%
%

\clearpage

%
%

\begin{figure}[h]
\begin	{center}
\leavevmode
\input	{plot_Eeffk2Aq0_n6f.tex}
\end	{center}
\caption{Effective energy of the k=2A, p=0, P=+ (variational) ground 
state of a flux tube of length $l/a=16,20,24,28,32,36,40,44,48,52,64$. 
Lines are our plateaux estimates ($\pm 1\sigma$ error bands).}
\label{fig_Eeffk2Aq0_n6f}
\end{figure}

\begin{figure}[htb]
\begin	{center}
\leavevmode
\input	{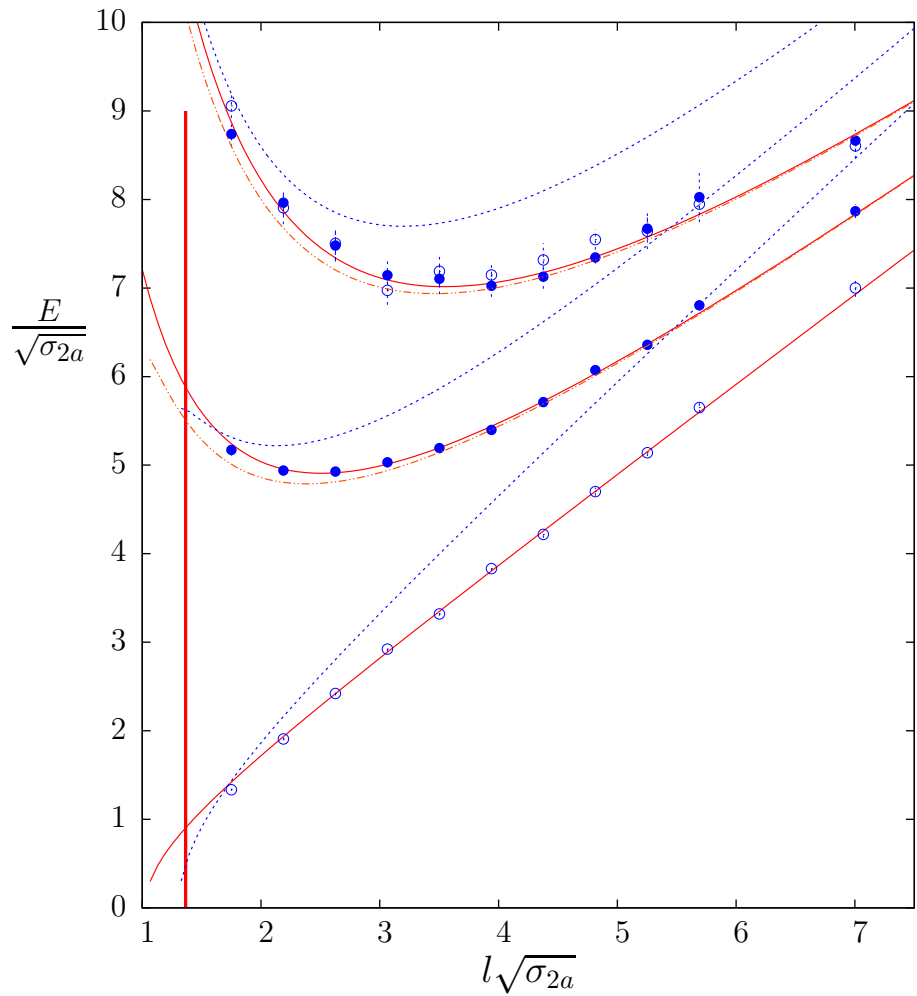}
\end	{center}
\caption{$k=2A$ ground states with $p=0,2\pi/l,4\pi/l$ and with $P=+$, 
$\circ$, and  $P=-$, $\bullet$. Solid red curves are Nambu-Goto predictions.
Dashed red lines are the model in eqn(\ref{eqn_Epmodel}). 
Dashed blue lines denotes lower boundaries of scattering states formed of two
fundamental flux tubes with total momentum $p$.
Vertical line denotes location of `deconfinement' transition.}
\label{fig_EgsQallk2a_n6f}
\end{figure}

\begin{figure}[htb]
\begin	{center}
\leavevmode
\input	{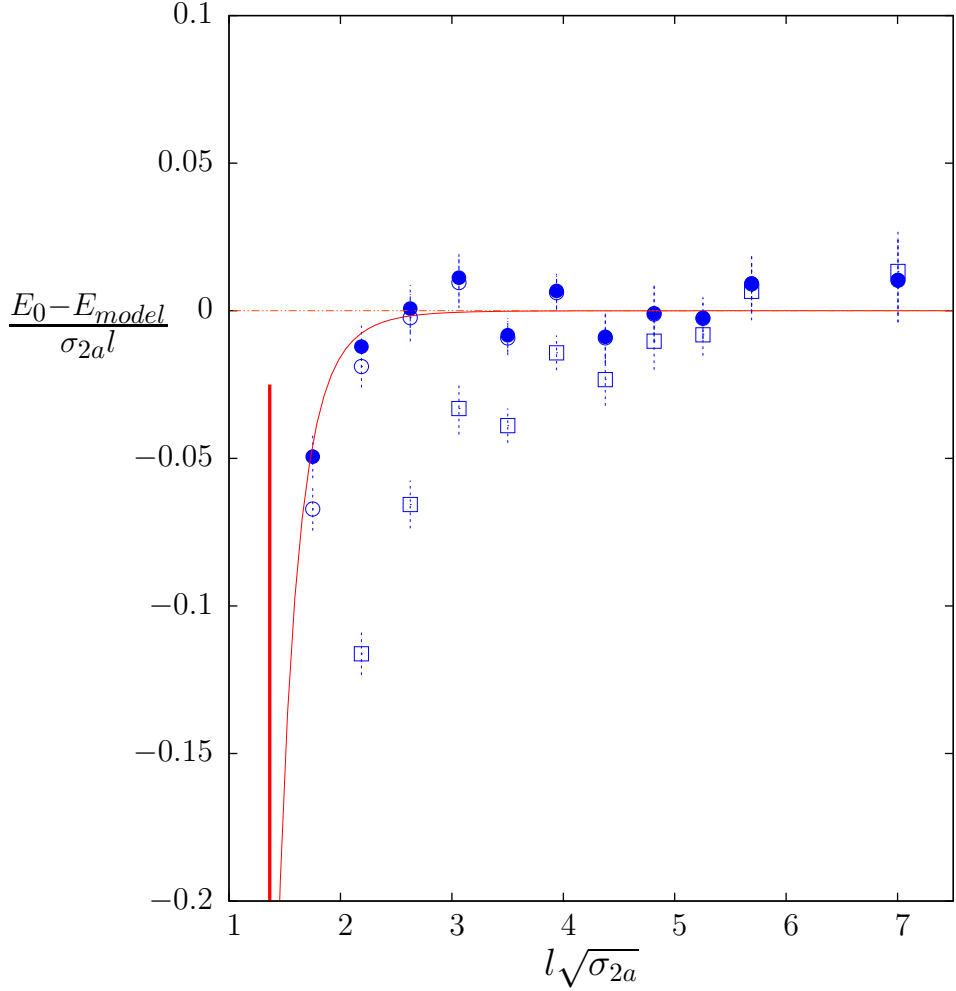}
\end	{center}
\caption{Energy of $k=2A$ ground state with $p=0$ and $P=+$, minus 
predictions of various `models': Nambu-Goto, $\bullet$; linear plus L\"uscher
correction, $\circ$; and only linear term, $\Box$. The solid curve
includes an $O(1/l^7)$ correction to Nambu-Goto.
Vertical line denotes location of `deconfinement' transition.}
\label{fig_DENGgsq0P+k2a_n6f}
\end{figure}

\begin{figure}[htb]
\begin	{center}
\leavevmode
\input	{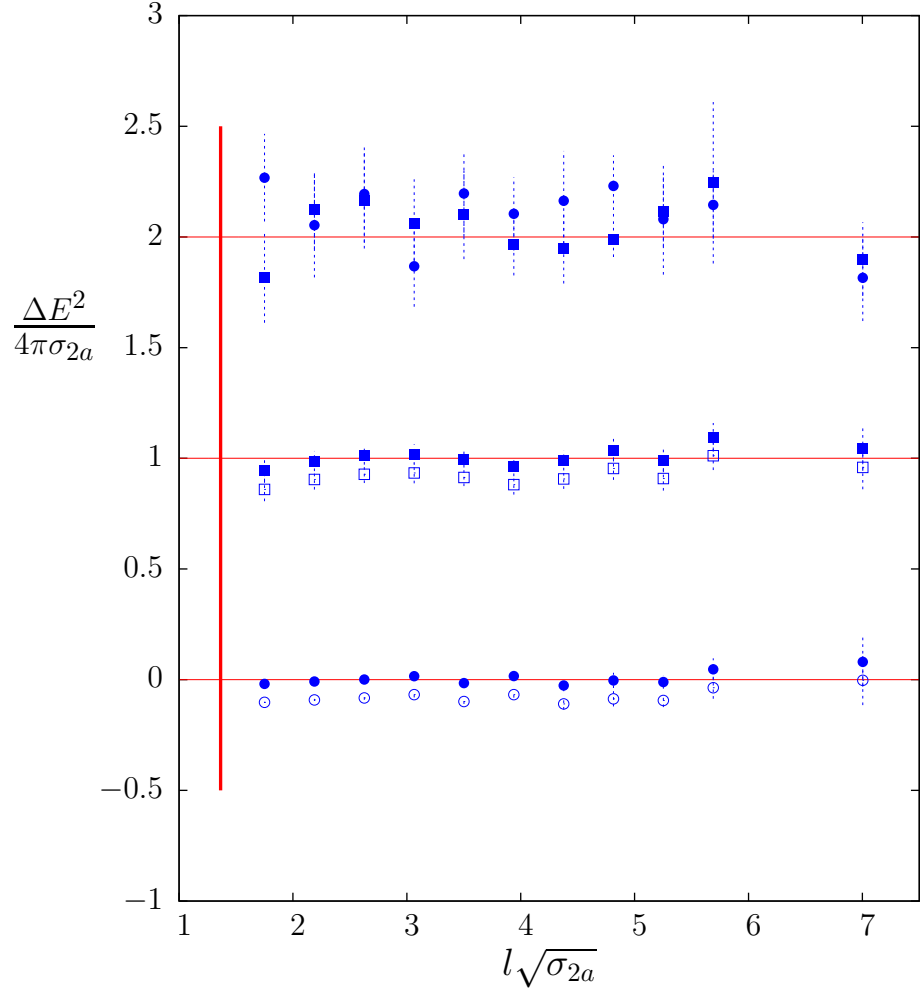}
\end	{center}
\caption{Phonon excitation energies, as defined in eqn(\ref{eqn_exq}), 
of $k=2A$ ground states with  $p=0, 2\pi/l, 4\pi/l$ 
and with  $P=+$ ($\bullet$) or $P=-$ ($\blacksquare$).
Open symbols shown for $p=0, 2\pi/l$ are without the zero-point 
energy in eqn(\ref{eqn_exq}).
Horizontal lines are Nambu-Goto  predictions.
Vertical line denotes location of `deconfinement' transition.}
\label{fig_DE2NGgsq0012k2a_n6f}
\end{figure}

\begin{figure}[htb]
\begin	{center}
\leavevmode
\input	{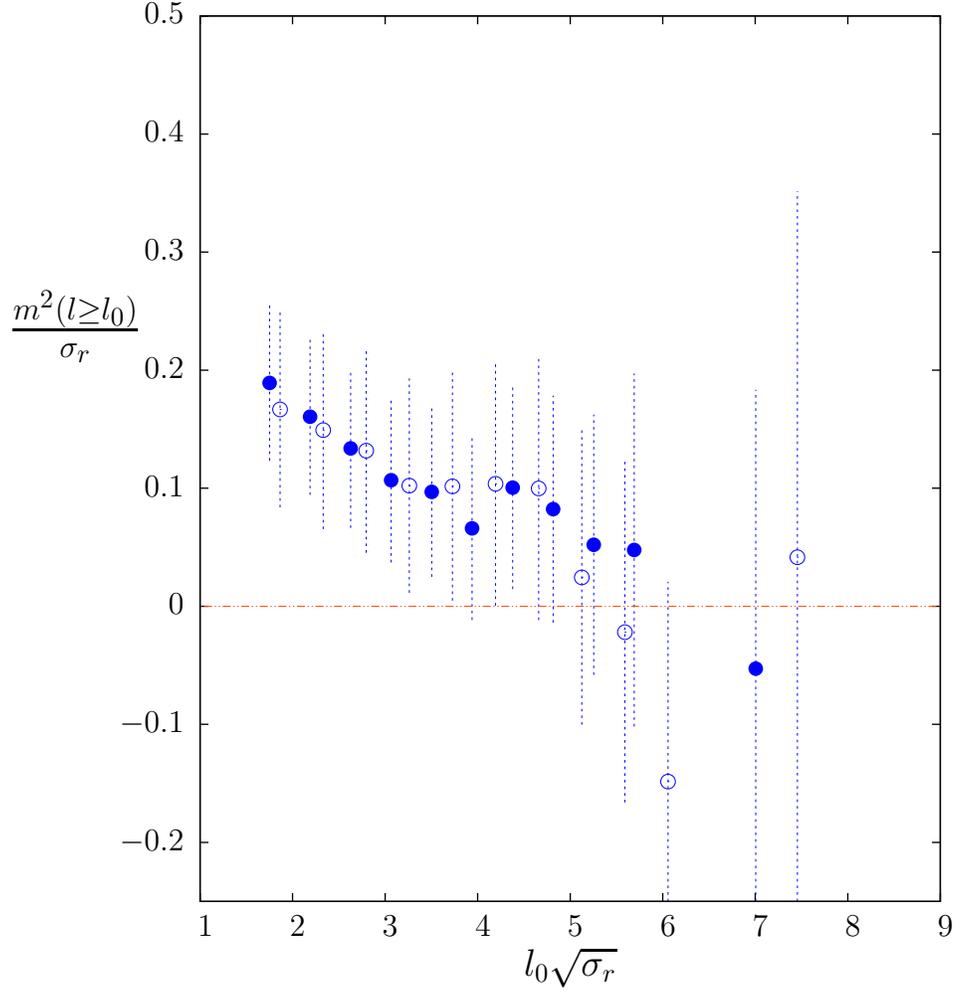}
\end	{center}
\caption{Fitting the  $p=2\pi/l$ ground state energies to the 
model in eqn(\ref{eqn_Epmodel}) and extracting the excitation mass averaged
over $l\geq l_0$. For representations $r=2A$ ($\bullet$) 
and  $r=3A$ ($\circ$).}
\label{fig_muq1k2Ak3A_n6f}
\end{figure}

\begin{figure}[htb]
\begin	{center}
\leavevmode
\input	{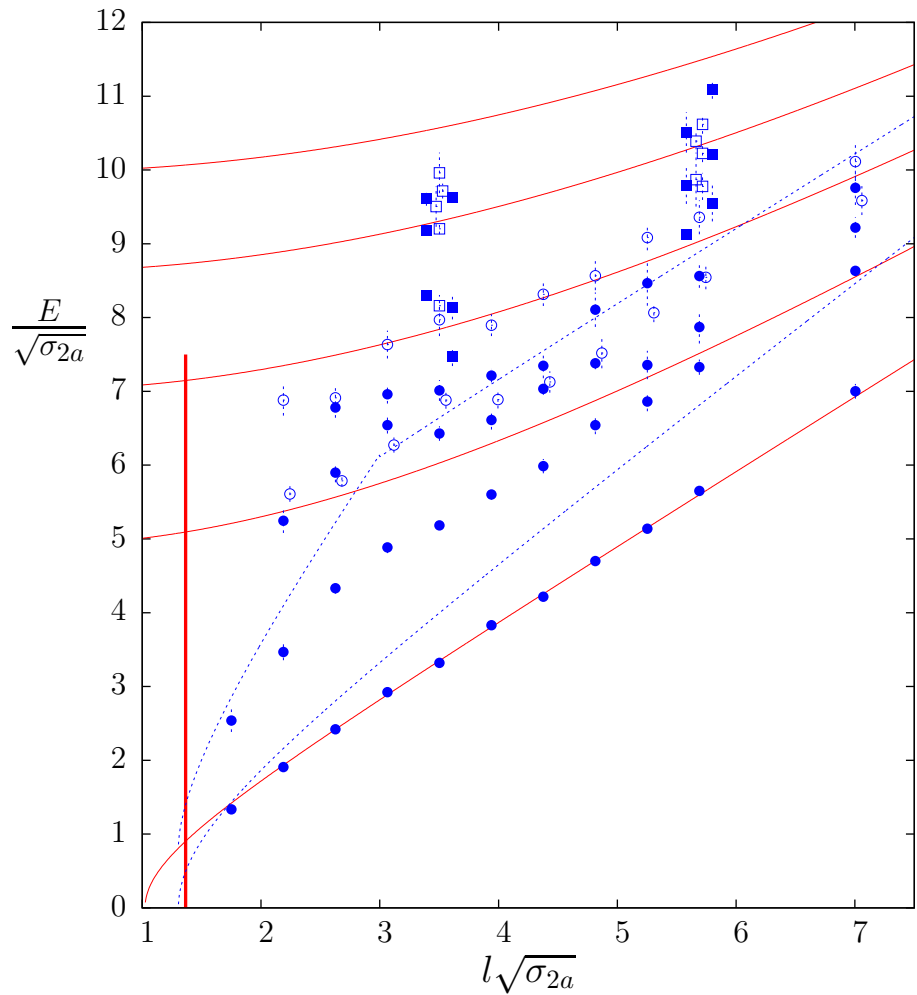}
\end	{center}
\caption{Energies of lightest $k=2A$  $p=0$ states with $P=+$, $\bullet$, and 
with $P=-$, $\circ$.  Solid curves are corresponding
Nambu-Goto levels. Upper dashed  line is (approximately) the energy of the
lightest $P=+$ decay channel consisting of a flux tube and  glueball; lower line
is that of two fundamental flux loops. 
Vertical line denotes location of `deconfinement' transition.}
\label{fig_Eq0k2a_n6f}
\end{figure}

\begin{figure}[h]
\begin	{center}
\leavevmode
\input	{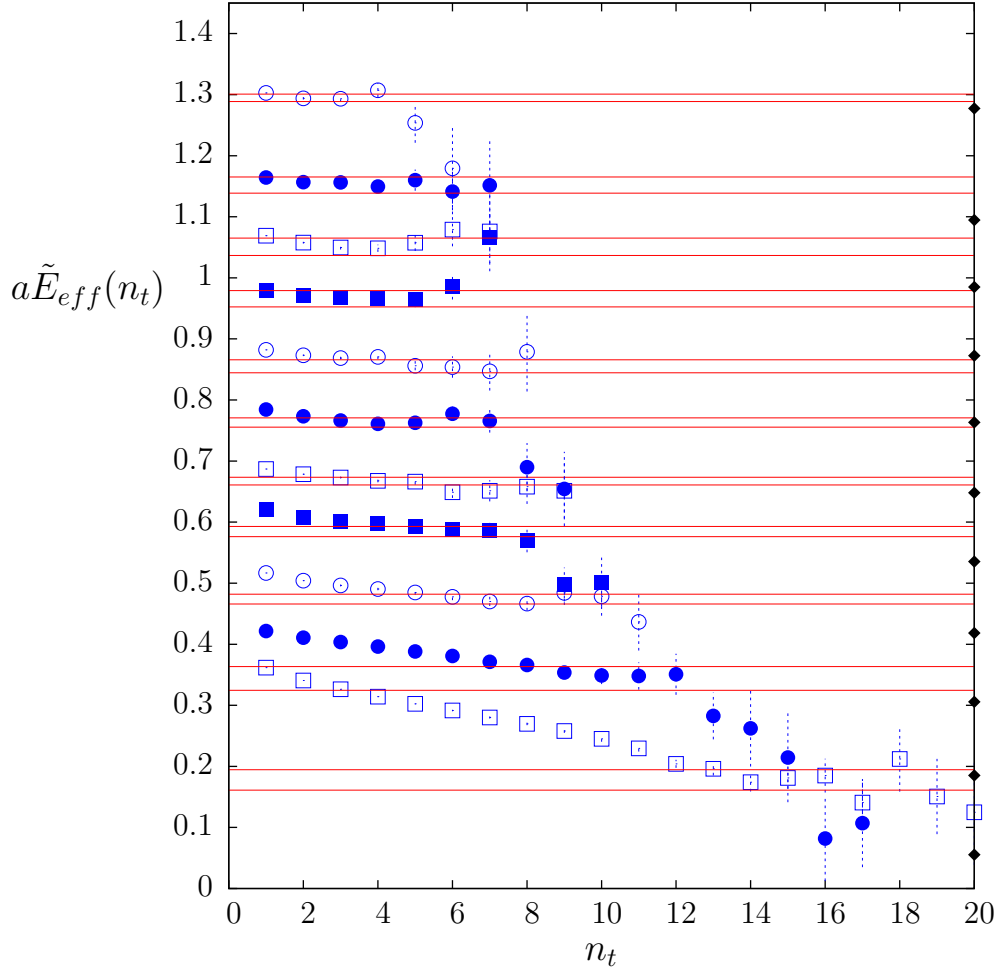}
\end	{center}
\caption{Effective energy, $E_{eff}$,  of the k=2A, p=0, P=+ (variational) first excited
state of a flux tube of length $l/a=16,20,24,28,32,36,40,44,48,52,64$. 
Values shown have been shifted by multiples of 0.05 for clarity, $a{\tilde{E}_{eff}} =
aE_{eff}+0.05*(l/a-24)/4$, and a shift of $0.35$ for $l=64a$ .  
Lines are $\pm 1\sigma$ error bands of our mean plateaux 
estimates. Points on right axis are corresponding decay thresholds ($=2E_f(l)$).}
\label{fig_Eeffk2Aq0ex_n6f}
\end{figure}

\begin{figure}[h]
\begin	{center}
\leavevmode
\input	{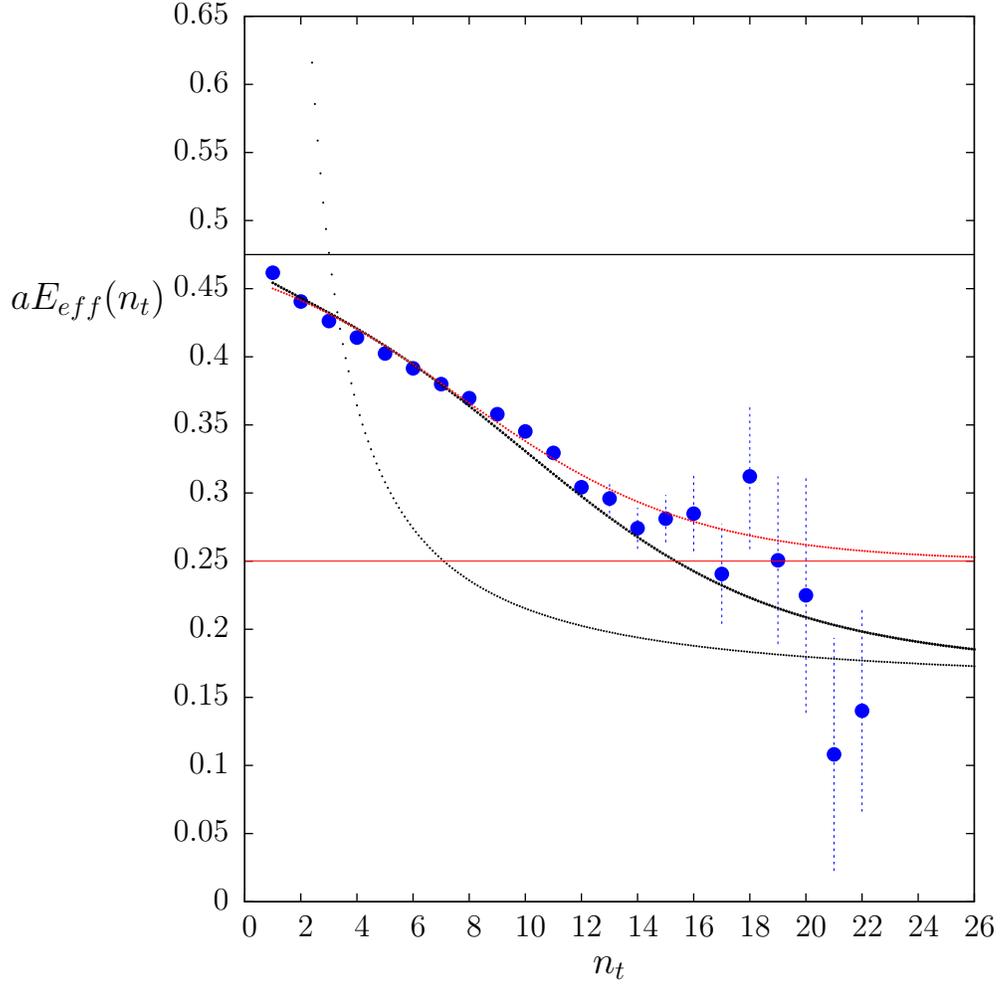}
\end	{center}
\caption{Effective energy, $E_{eff}$,  of the k=2A, p=0, P=+ (variational) first excited
state of a flux tube of length $l=16a$. The red curve is what one obtains
with an excited state, $aE^\star=0.49$, in addition to the ground
state, which is indicated by the lower horizontal line. 
The solid black curve is obtained by summing over scattering states of 2
fundamental flux tubes with a Breit-Wigner amplitude peaking at
the upper horizontal line and with a width  $a\Gamma=0.065$. The lower black
curve sums uniformly over all scattering states.}  
\label{fig_Eeffk2Aq0ex_l16_n6f}
\end{figure}

\clearpage
%
%

\begin{figure}[htb]
\begin	{center}
\leavevmode
\input	{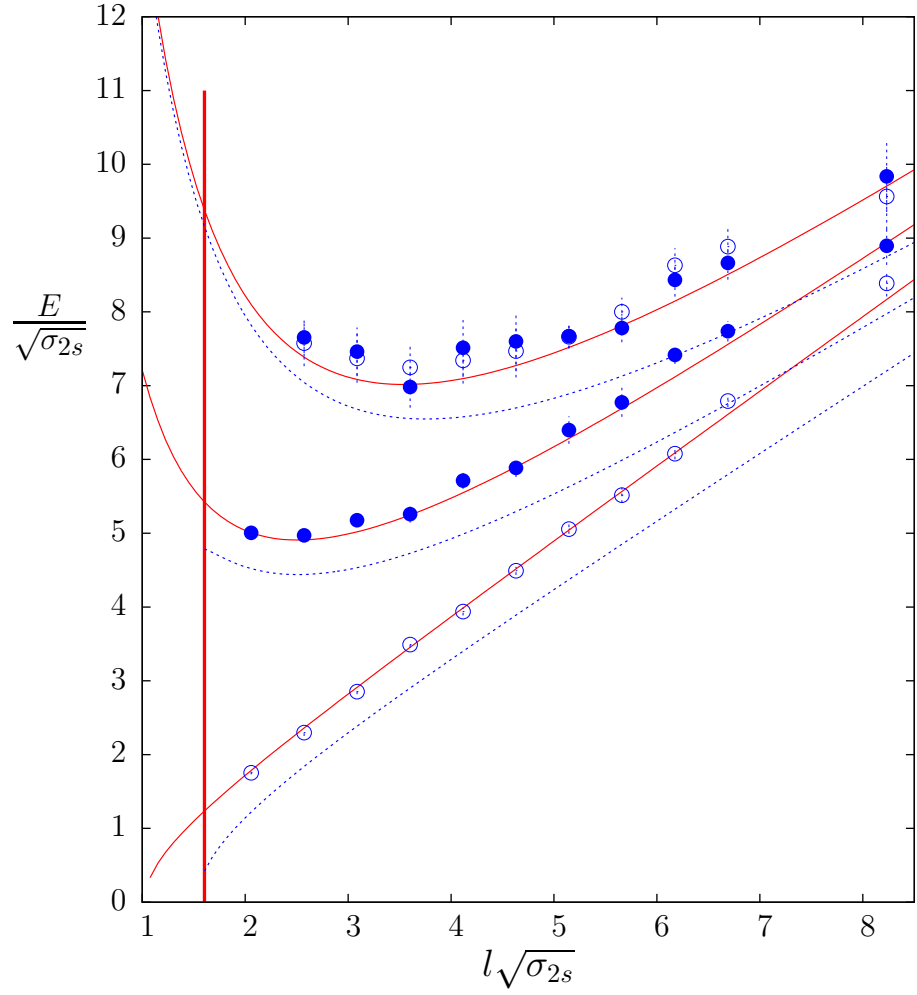}
\end	{center}
\caption{$k=2S$ ground states with $p=0,2\pi/l,4\pi/l$ and with $P=+$, 
$\circ$, and  $P=-$, $\bullet$. Curves are Nambu-Goto predictions. 
Dashed lines are thresholds for scattering states formed of two
fundamental flux tubes with total momentum $p$.
Vertical line denotes location of `deconfinement' transition.}
\label{fig_EgsQallk2s_n6f}
\end{figure}

\begin{figure}[htb]
\begin	{center}
\leavevmode
\input	{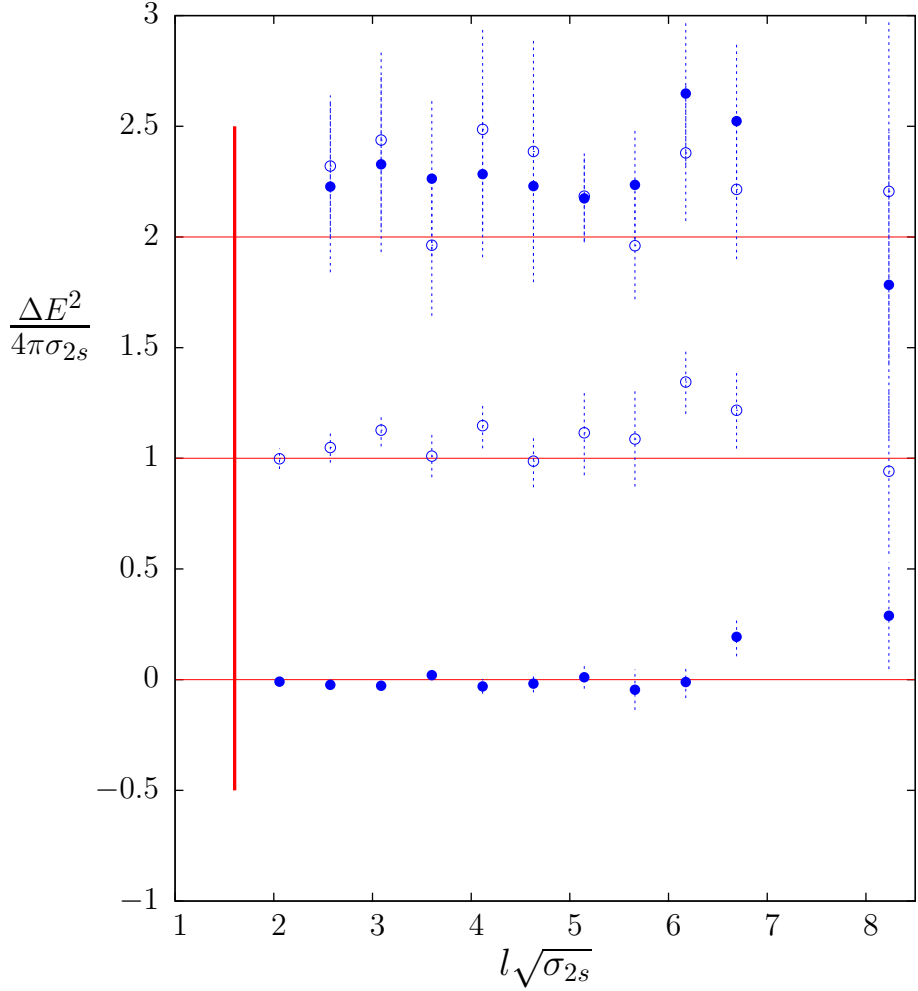}
\end	{center}
\caption{Phonon excitation energies (see eqn(\ref{eqn_exq})) 
of lowest $k=2S$ states with $p=0$ and
  $P=+$, $\bullet$,  $p=2\pi/l$ and $P=-$, $\circ$,  
and with $p=4\pi/l$ for both $P=-$, $\circ$ and $P=+$, $\bullet$.
Horizontal lines are Nambu-Goto  predictions.
Vertical line denotes location of `deconfinement' transition.}
\label{fig_DE2NGgsq0012k2s_n6f}
\end{figure}

\begin{figure}[h]
\begin	{center}
\leavevmode
\input	{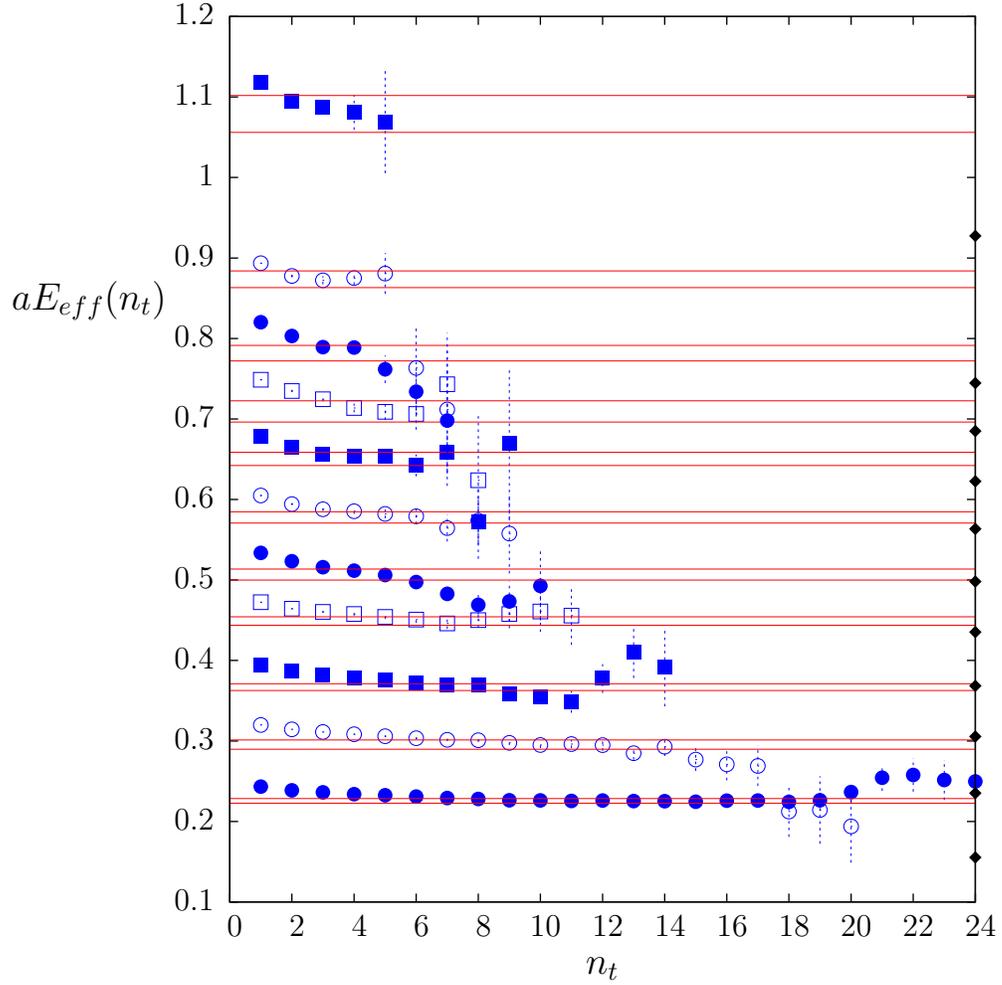}
\end	{center}
\caption{Effective energy of the k=2S, p=0, P=+ (variational) ground 
state of a flux tube of length $l/a=16,20,24,28,32,36,40,44,48,52,64$.
Plateau estimate ($\pm 1\sigma$ error band) given by red lines. Lowest energy of 
two fundamental lines is indicated by diamonds on right axis. }
\label{fig_Eeffk2Sq0_n6f}
\end{figure}

\begin{figure}[h]
\begin	{center}
\leavevmode
\input	{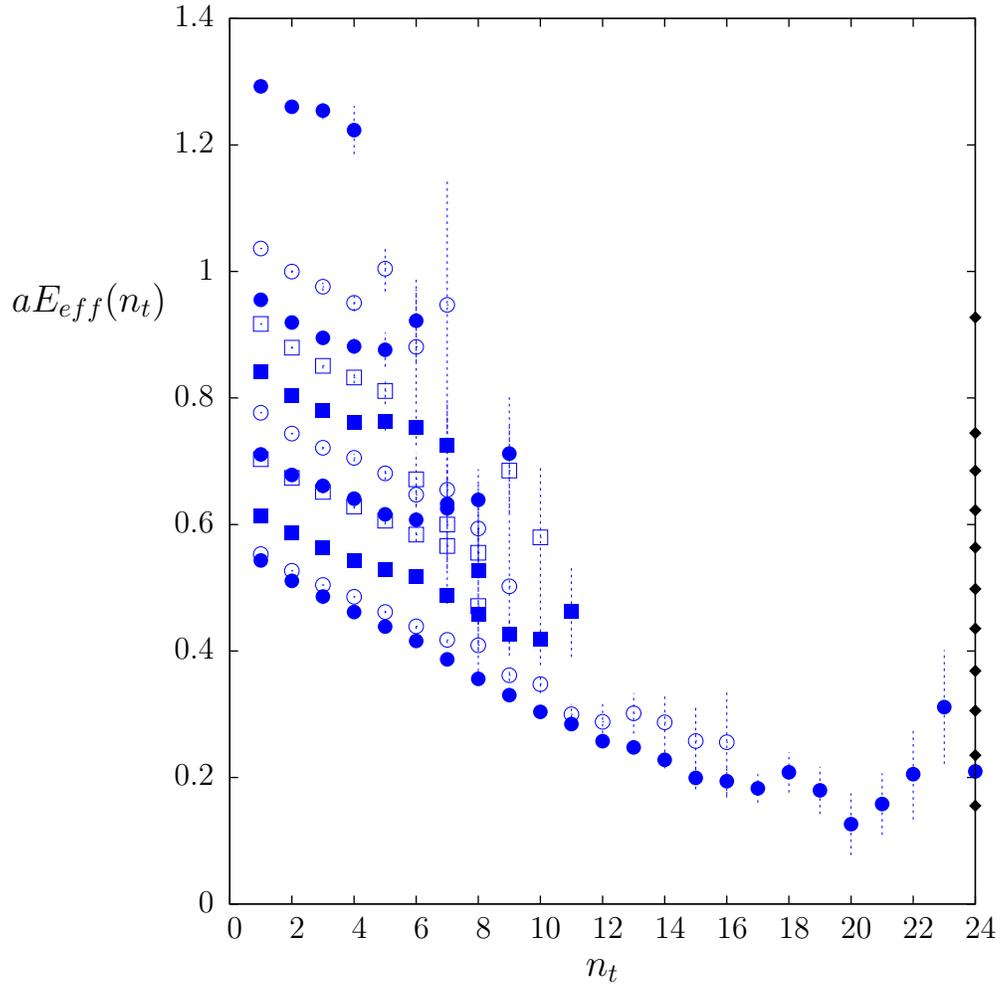}
\end	{center}
\caption{Effective energy of the k=2S, p=0, P=+ (variational) first excited
state of a flux tube of length $l/a=16,20,24,28,32,36,40,44,48,52,64$.
Lowest energy of two fundamental lines is indicated by diamonds
on right vertical axis. }
\label{fig_Eeffk2Sq0ex_n6f}
\end{figure}

\begin{figure}[h]
\begin	{center}
\leavevmode
\input	{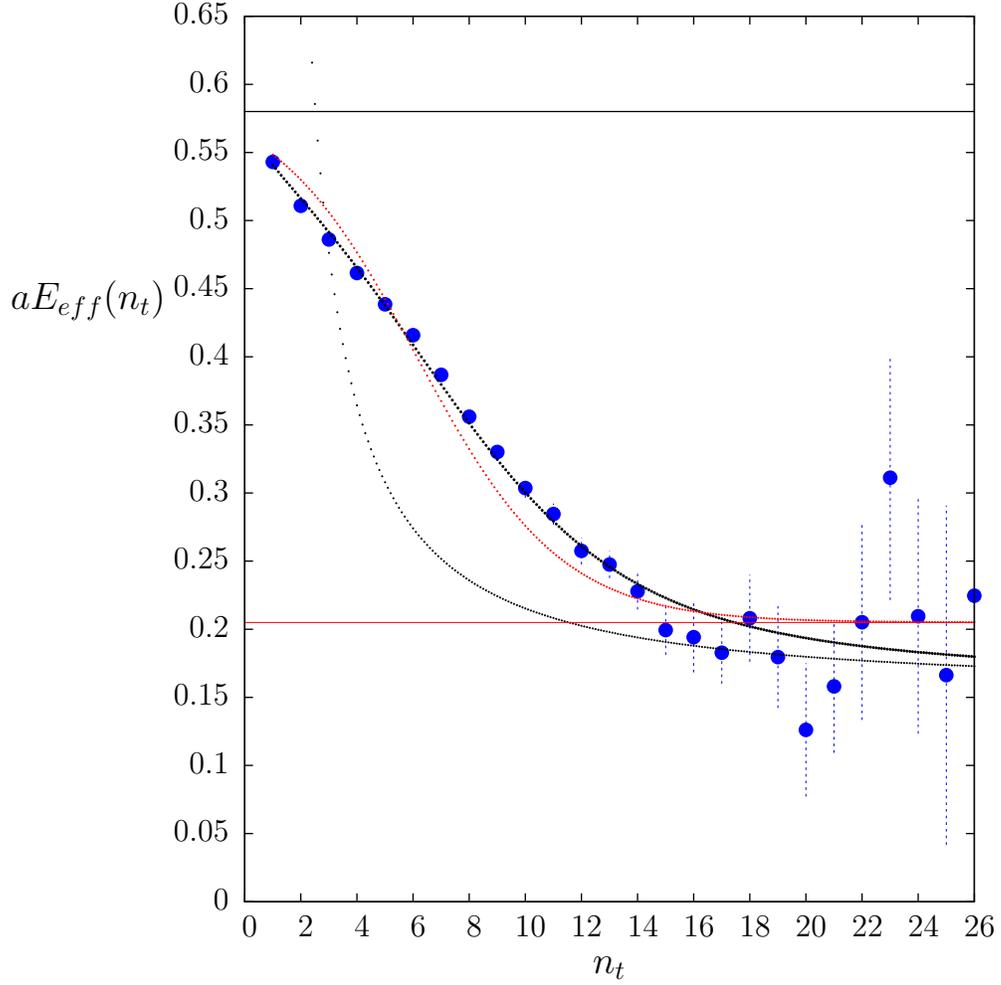}
\end	{center}
\caption{Effective energy, $E_{eff}$,  of the k=2S, p=0, P=+ (variational) first excited
state of a flux tube of length $l=16a$. The  red curve is an example of fitting
with an excited state, $aE^\star=0.595$, in addition to the ground
state indicated by the lower horizontal solid line. 
The upper black curve is obtained by summing over scattering states of 2
fundamental flux tubes with a Breit-Wigner probability peaking at
the upper horizontal line, with width $a\Gamma=0.12$. The lower black
curve sums uniformly over all scattering states.}
\label{fig_Eeffk2Sq0ex_l16_n6f}
\end{figure}

\clearpage

%
%

\begin{figure}[h]
\begin	{center}
\leavevmode
\input	{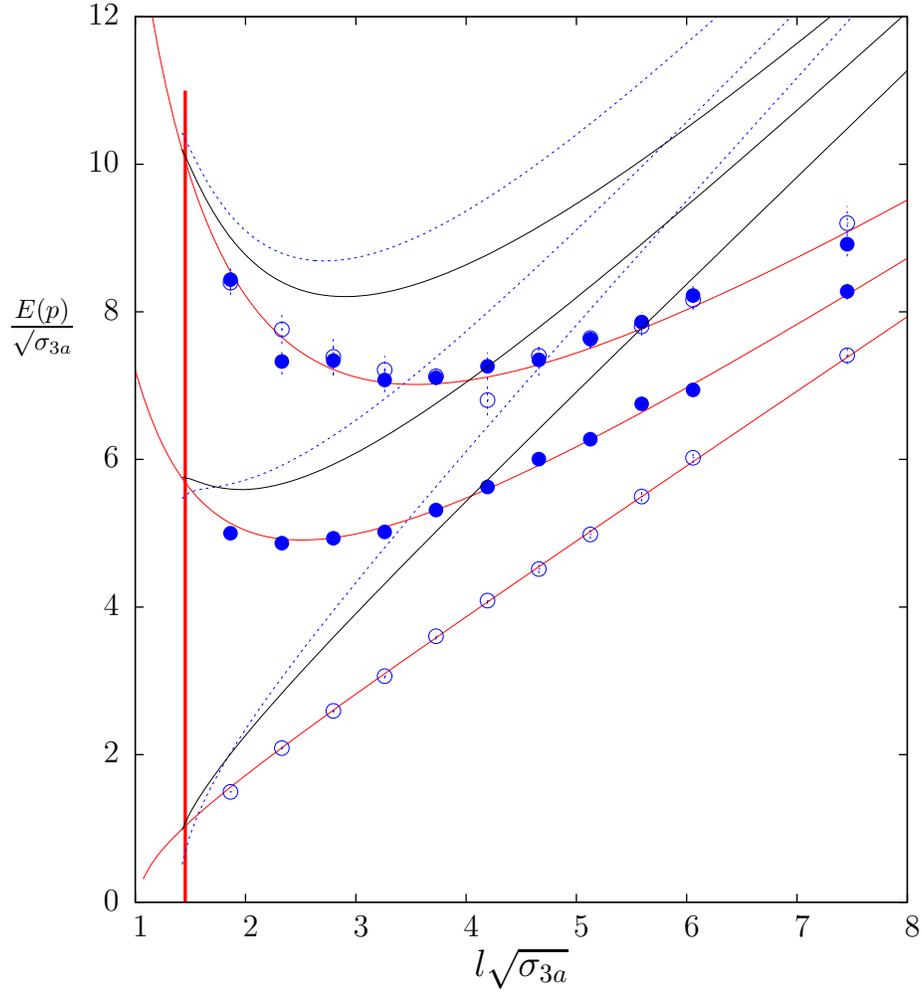}
\end	{center}
\caption{$k=3A$ ground states with $p=0,2\pi/l,4\pi/l$ and with $P=+$, 
$\circ$, and $P=-$, $\bullet$. Red curves are Nambu-Goto predictions. 
Black dashed line denotes lower boundary of scattering state formed of three
fundamental flux tubes with total momentum $p$, and black solid line of a 
$k=2A$ flux tube with a fundamental flux tube with total momentum $p$.
Vertical line denotes location of `deconfinement' transition.}
\label{fig_EQgsk3A_n6f}
\end{figure}

\begin{figure}[h]
\begin	{center}
\leavevmode
\input	{plot_Eeffk3Aq0_n6f.tex}
\end	{center}
\caption{Effective energy of the k=3A, p=0, P=+ (variational) ground 
state of a flux tube of length $l/a=16,20,24,28,32,36,40,44,48,52,64$. 
Lines are our plateaux estimates ($\pm 1\sigma$ error bands).}
\label{fig_Eeffk3Aq0_n6f}
\end{figure}

\begin{figure}[h]
\begin	{center}
\leavevmode
\input	{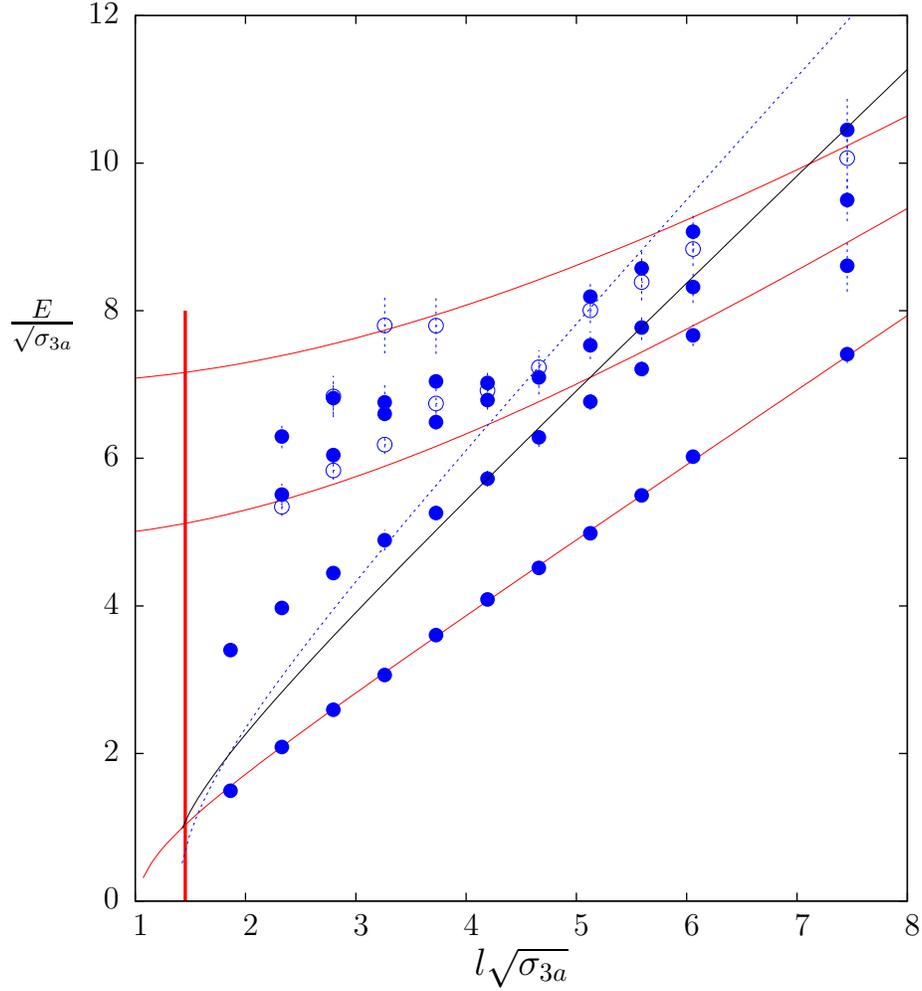}
\end	{center}
\caption{Energies of lightest $k=3A$ states with $p=0$ and with $P=+$, 
$\bullet$, or $P=-$, $\circ$. Red curves are corresponding
Nambu-Goto levels. Black dashed line is the lowest energy 
of three fundamental flux loops; solid black line is that of a
$k=2A$ flux tube  with a fundamental one.
Vertical line denotes location of `deconfinement' transition.}
\label{fig_EQ0k3A_n6f}
\end{figure}

\begin{figure}[h]
\begin	{center}
\leavevmode
\input	{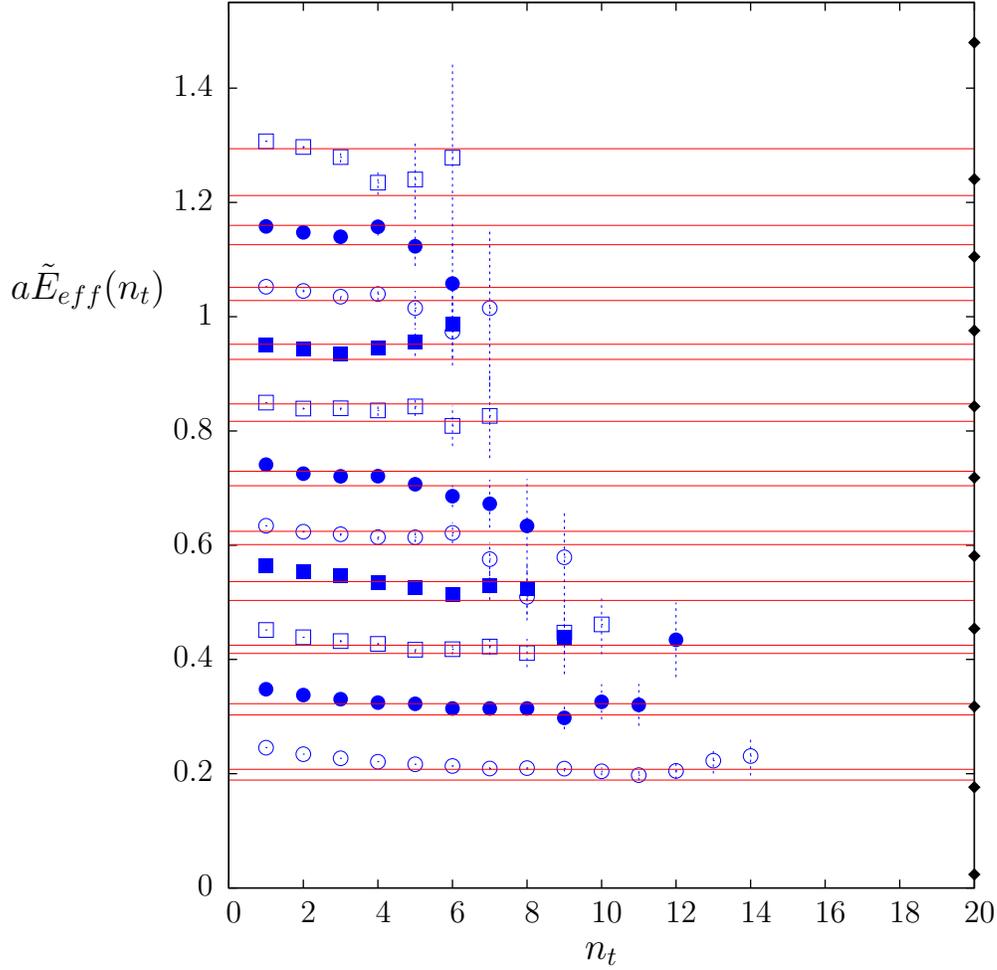}
\end	{center}
\caption{Effective energy, $aE_{eff}$, of the k=3A, p=0, P=+ (variational) first 
excited state of a flux tube of length $l/a=16,20,24,28,32,36,40,44,48,52,64$. 
Values shown have been shifted for clarity: $a{\tilde{E}_{eff}} =
aE_{eff}+0.05*(l/a-32)/4$. (Shift of $l=64a$ is $0.25$.) Lines are our plateau estimates 
($\pm 1\sigma$ error bands).
Thresholds for decay shown as diamonds on right axis.}
\label{fig_Eeffk3Aq0ex_n6f}
\end{figure}

\begin{figure}[h]
\begin	{center}
\leavevmode
\input	{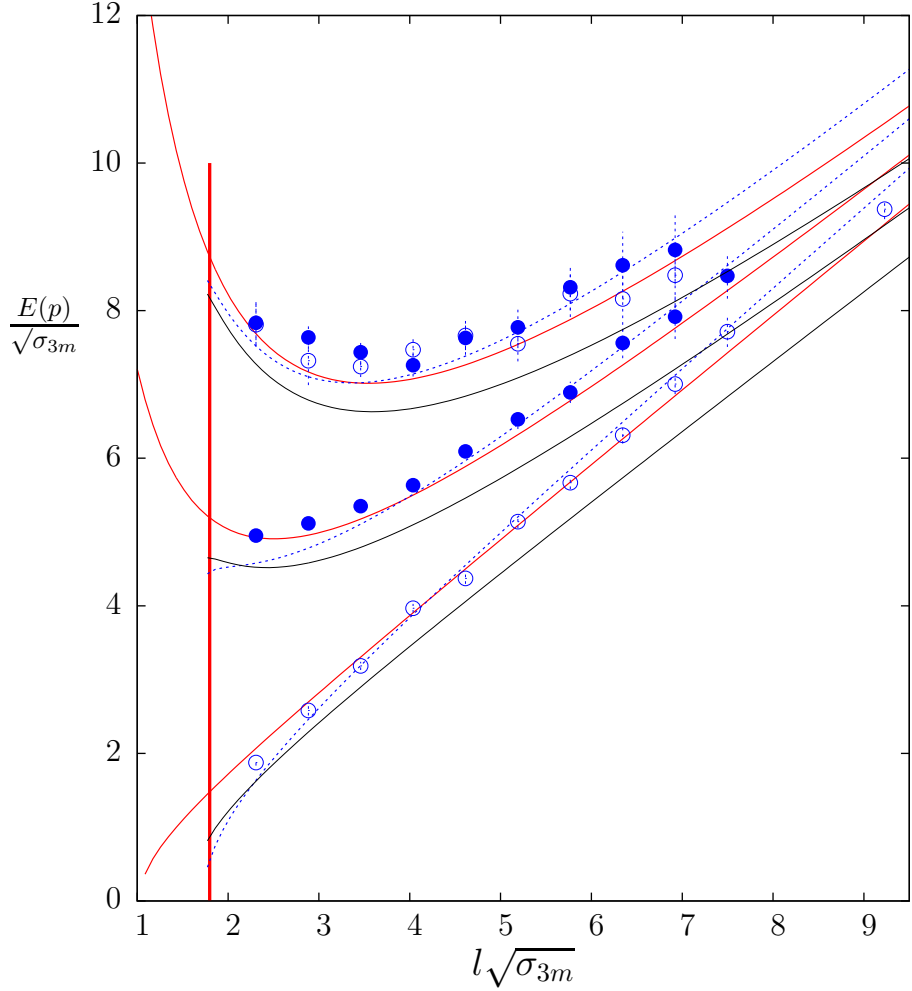}
\end	{center}
\caption{$k=3M$ ground states with $p=0,2\pi/l,4\pi/l$ and with $P=+$, 
$\circ$, and $P=-$, $\bullet$. Solid red curves are Nambu-Goto predictions. 
Dashed line denotes lower boundary of scattering state formed of three
fundamental flux tubes with total momentum $p$; black line of a $k=2A$
flux tube with a fundamental flux tube.
Vertical line denotes location of `deconfinement' transition.}
\label{fig_EQgsk3M_n6f}
\end{figure}

\begin{figure}[h]
\begin	{center}
\leavevmode
\input	{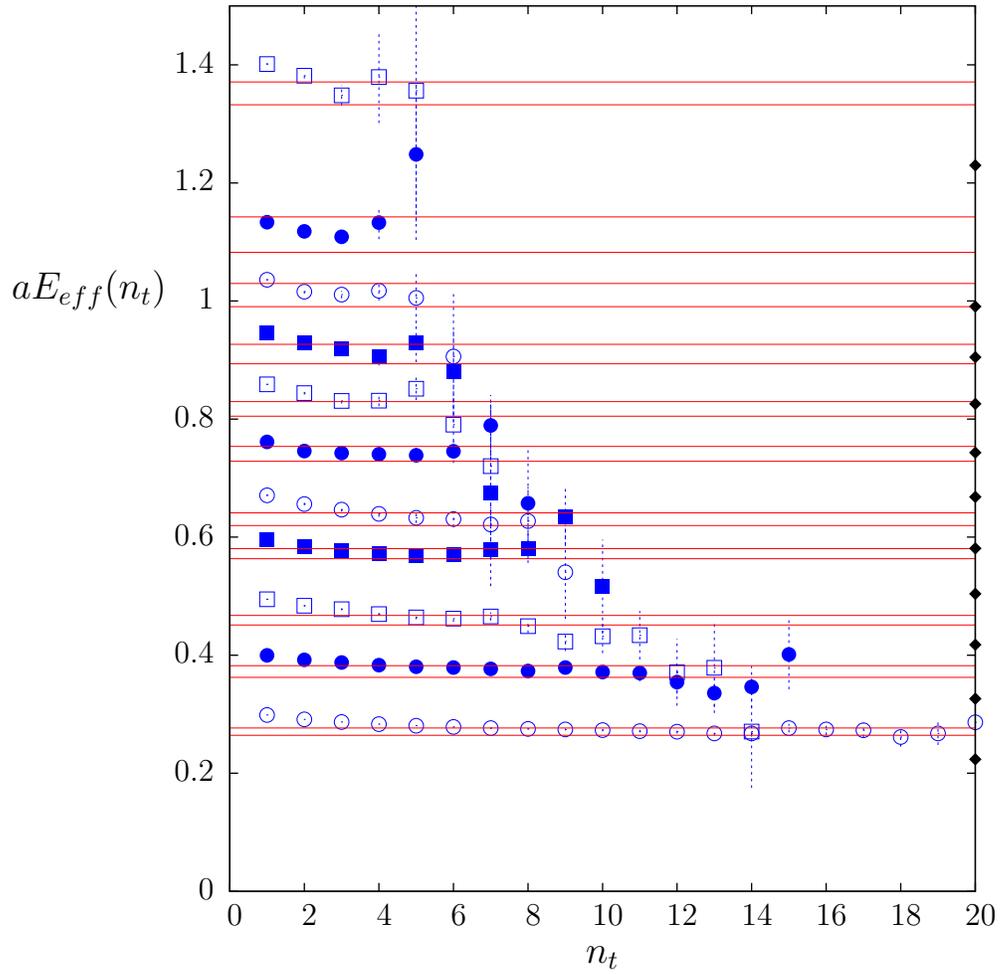}
\end	{center}
\caption{Effective energy of the k=3M, p=0, P=+ (variational) ground 
state of a flux tube of length $l/a=16,20,24,28,32,36,40,44,48,52,64$. 
Lines are our plateaux estimates ($\pm 1\sigma$ error bands).
Decay thresholds indicated by diamonds on right axis.}
\label{fig_Eeffk3Mq0_n6f}
\end{figure}

\begin{figure}[h]
\begin	{center}
\leavevmode
\input	{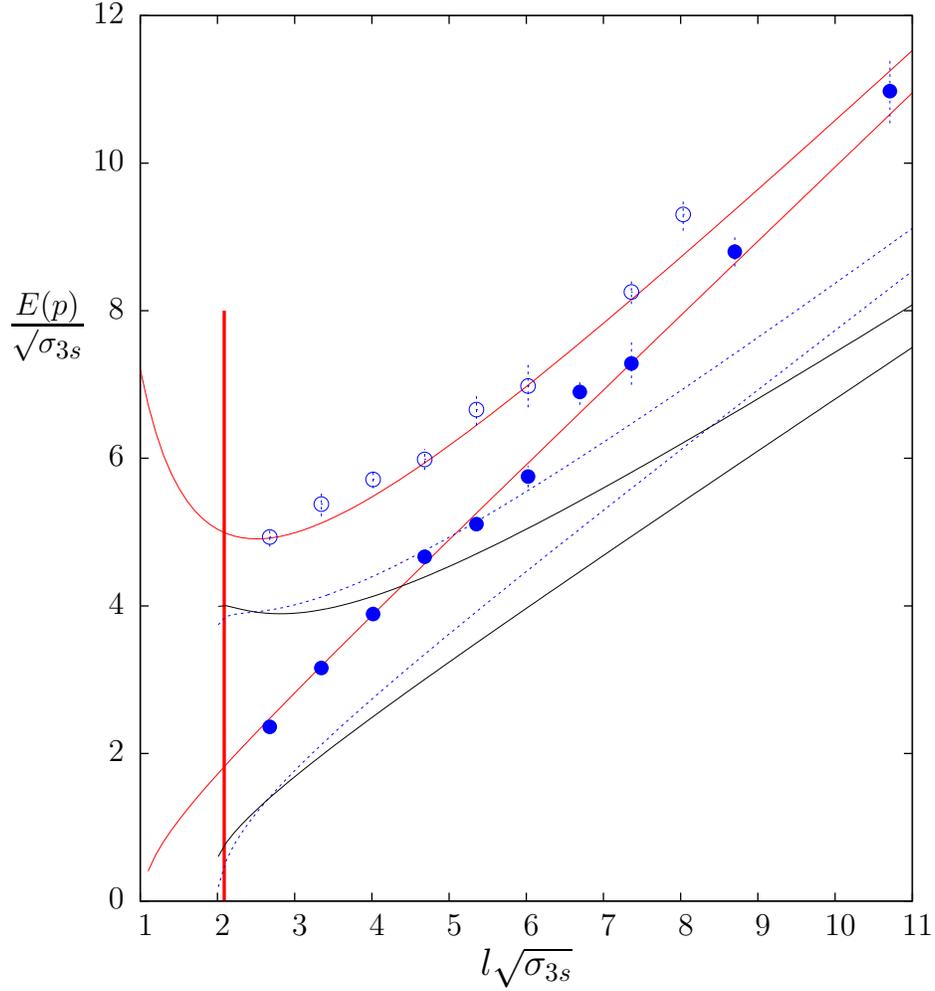}
\end	{center}
\caption{$k=3S$ ground states with $p=0,P=+$, $\bullet$, and with 
$p=2\pi/l, P=-$, $\circ$. Solid red curves are Nambu-Goto predictions.
Dashed line denotes lower boundary of scattering state formed of three
fundamental flux tubes with same momentum; black line of a $k=2A$
flux tube with a fundamental flux tube.
Vertical line denotes location of `deconfinement' transition.}
\label{fig_EQgsk3S_n6f}
\end{figure}

\begin{figure}[h]
\begin	{center}
\leavevmode
\input	{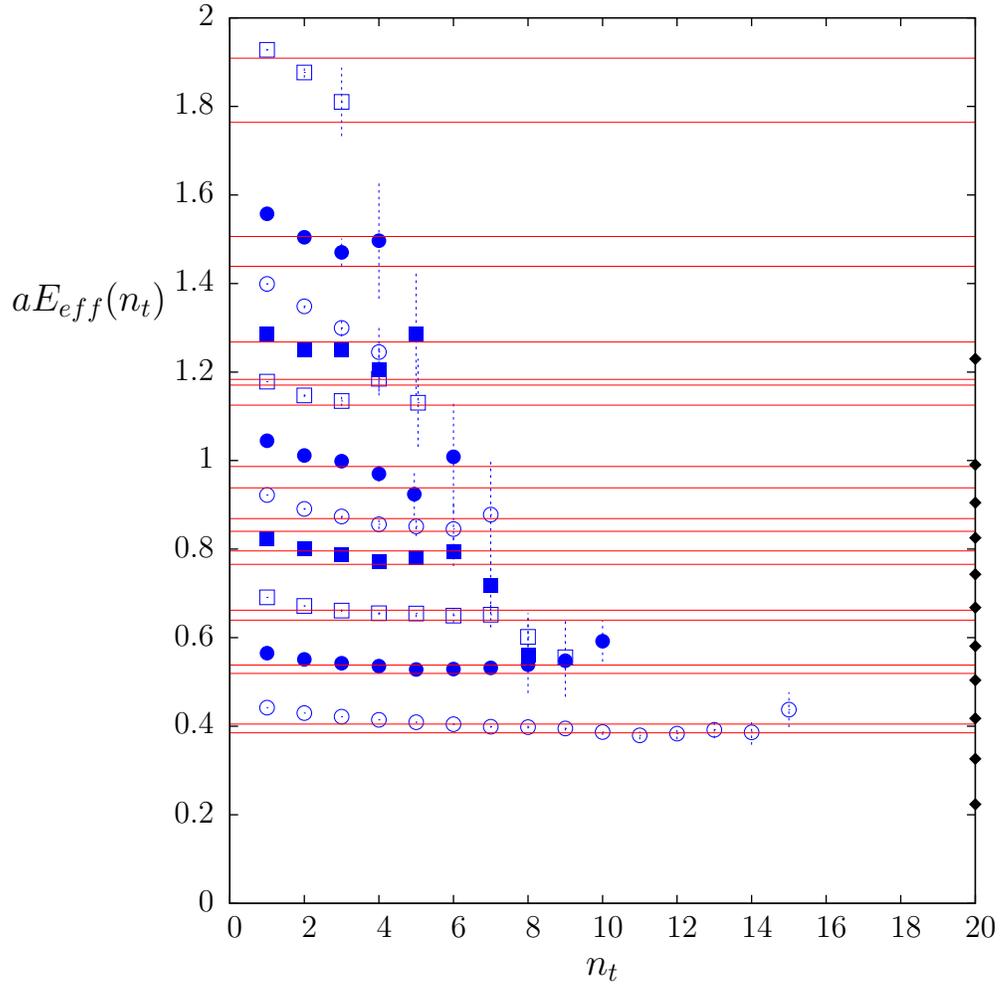}
\end	{center}
\caption{Effective energy of the k=3S, p=0, P=+ (variational) ground 
state of a flux tube of length $l/a=16,20,24,28,32,36,40,44,48,52,64$. 
Lines are our plateaux estimates ($\pm 1\sigma$ error bands). No plateau 
attempted for $l=48a$. Diamonds on right axis denote decay thresholds.}
\label{fig_Eeffk3Sq0n_n6f}
\end{figure}

\clearpage

%
%

\begin{figure}[htb]
\begin	{center}
\leavevmode
\input	{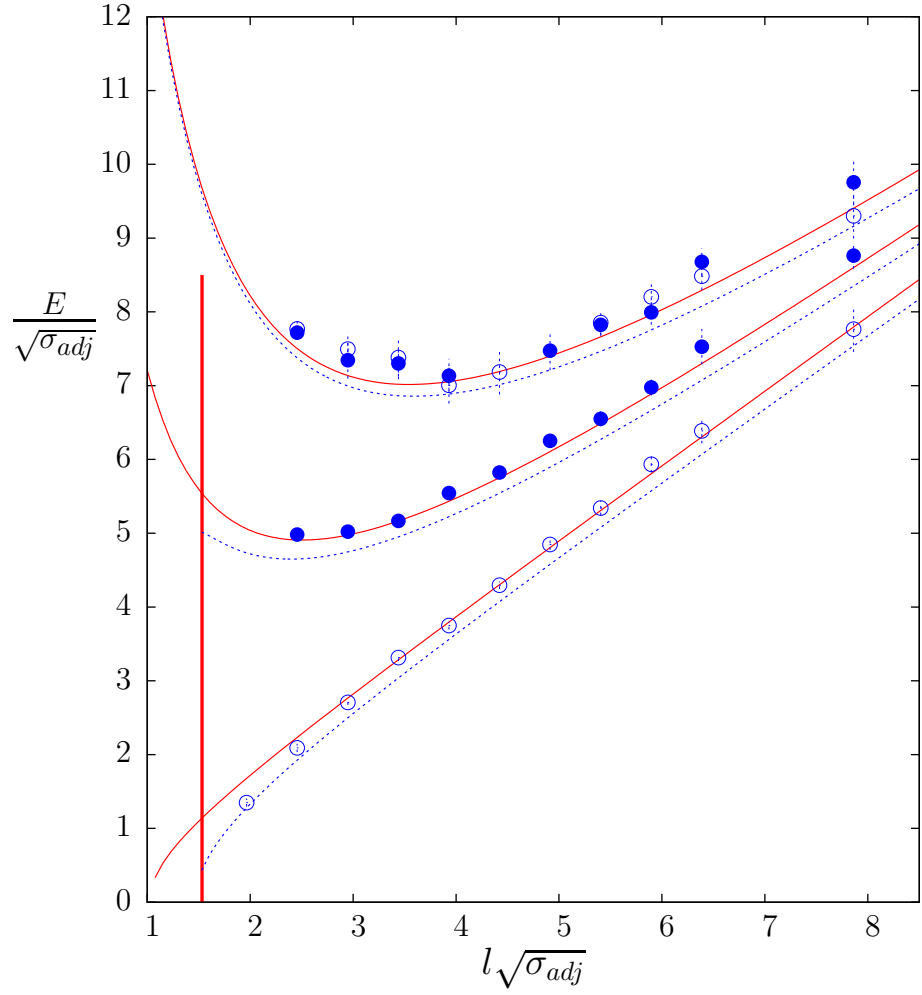}
\end	{center}
\caption{Adjoint ground states with $p=0,2\pi/l,4\pi/l$ and with $P=+$, 
$\circ$, and $P=-$, $\bullet$. Solid red curves are Nambu-Goto 
predictions. Dashed lines denote lower boundaries of scattering state 
formed of a pair of (anti)fundamental flux tubes with same momentum.
Vertical line denotes location of `deconfinement' transition.}
\label{fig_EgsQallAdj_n6f}
\end{figure}

\begin{figure}[h]
\begin	{center}
\leavevmode
\input	{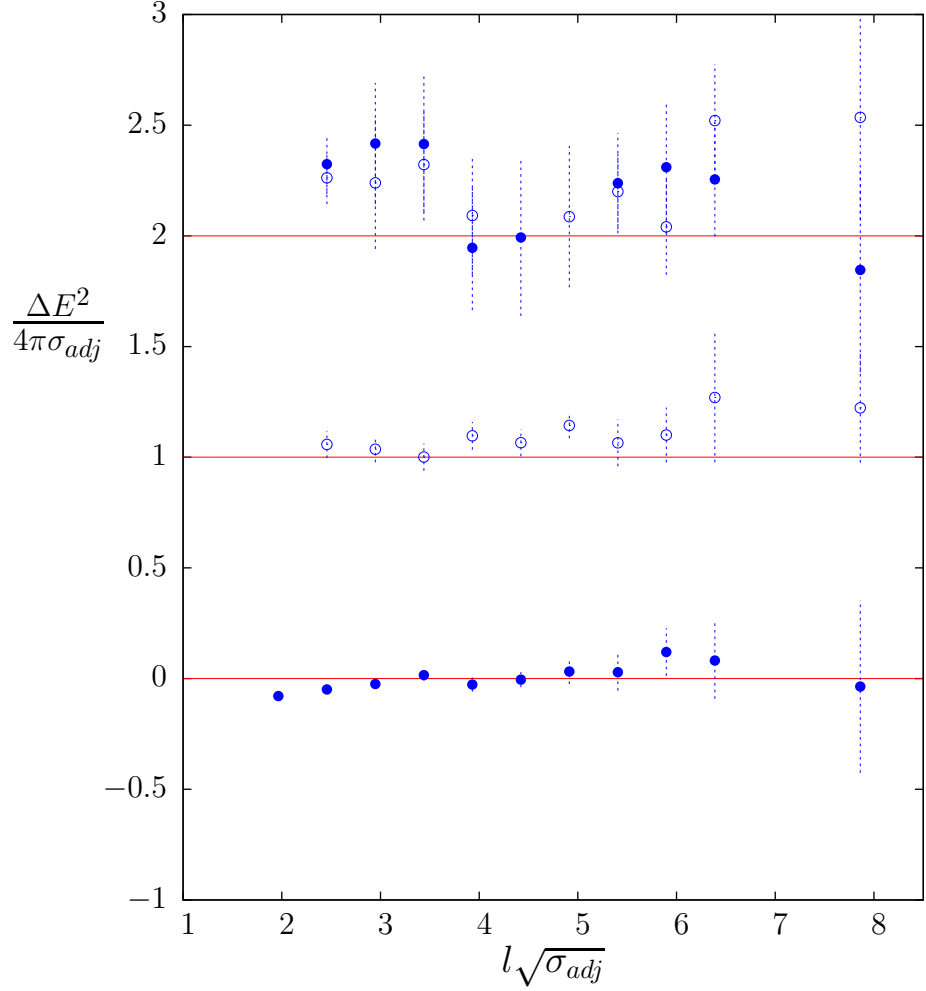}
\end	{center}
\caption{Phonon excitation energies of adjoint ground states with $p=0$ and
  $P=+$, $\bullet$,  $p=2\pi/l$ and $P=-$, $\circ$,  
and with $p=4\pi/l$ for both $P=-$, $\circ$ and $P=+$, $\bullet$.
Lines are Nambu-Goto  predictions.}
\label{fig_DE2NGgsq012adj_n6f} 
\end{figure}

\begin{figure}[htb]
\begin	{center}
\leavevmode
\input	{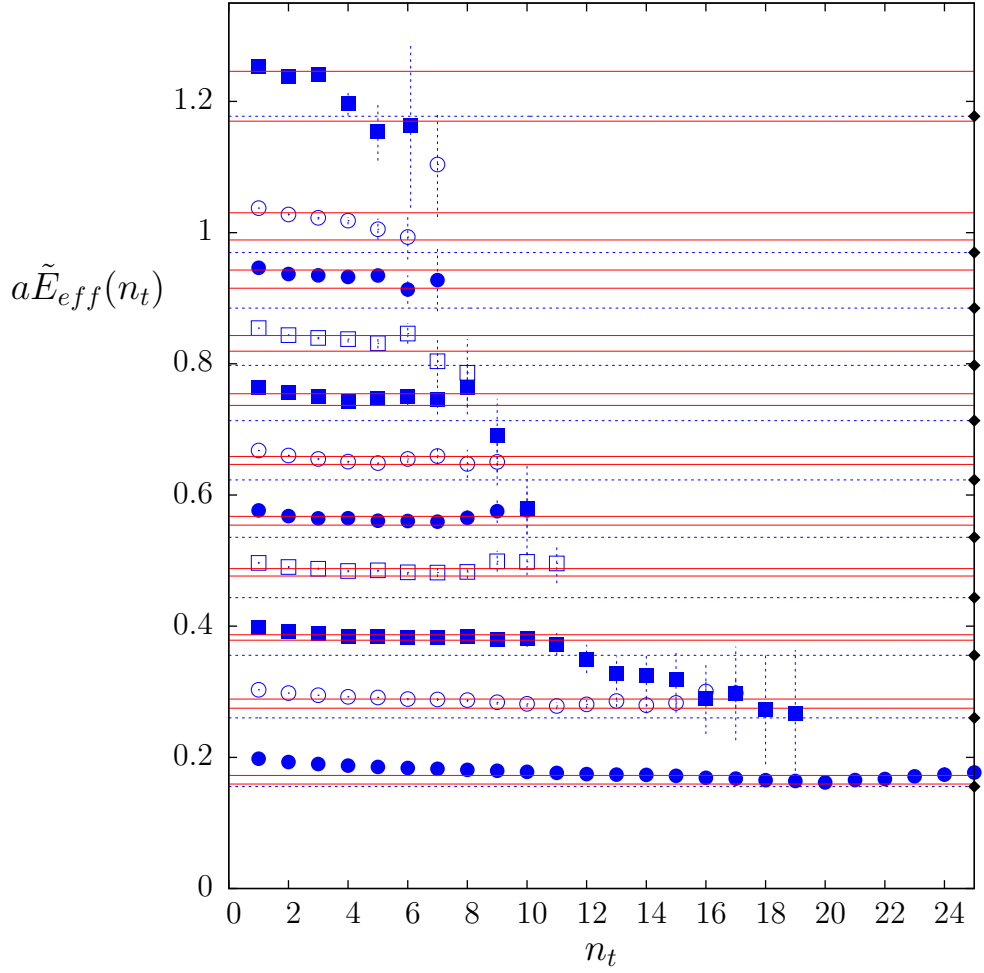}
\end	{center}
\caption{Ground state adjoint loop effective masses for $l/a=$16, 20, 24,
  28, 32, 36, 40, 44, 48, 52, 64.
Solid lines give $\pm 1\sigma$ error bands of our plateaux estimates. Dashed lines
are twice the energy of corresponding (anti)fundamental loops. (Also
indicated by diamonds on right axis.) Values shown
have been shifted by multiples of 0.025 for clarity: $a{\tilde{E}_{eff}} =
aE_{eff} +0.025*(l/a-16)/4$, except a shift of $0.25$ for $l=64a$. }
\label{fig_Eeffadjb_n6f}
\end{figure}

\begin{figure}[htb]
\begin	{center}
\leavevmode
\input	{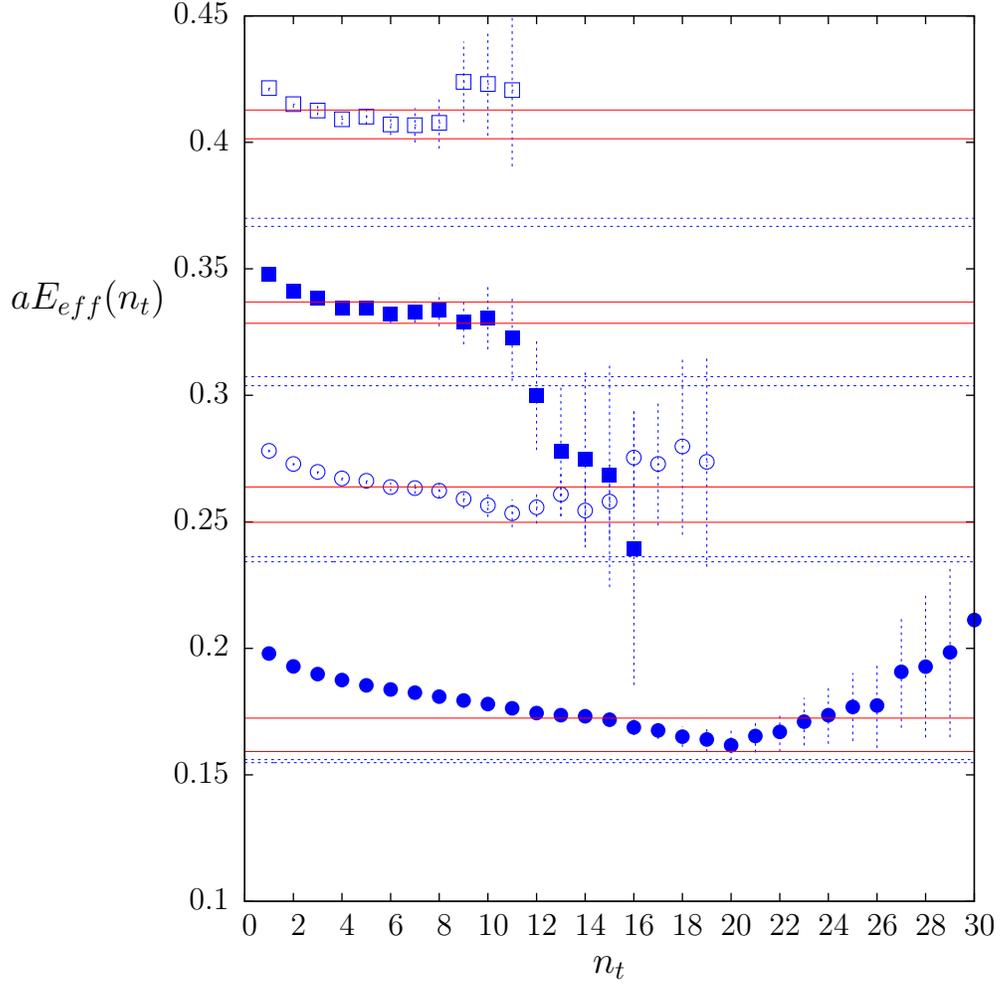}
\end	{center}
\caption{Ground state adjoint loop effective masses for $l/a=16,20,24,28$.
Solid lines give $\pm 1 \sigma$ error bands for our plateaux estimates. 
Dashed lines are error bands for twice the energy of corresponding 
(anti)fundamental loops.} 
\label{fig_Eeffadjlb_n6f}
\end{figure}

%
%

\begin{figure}[htb]
\begin	{center}
\leavevmode
\input	{plot_EgsQallr84_n6f.tex}
\end	{center}
\caption{Ground states in the $\underline{84}$ representation for 
$p=0,2\pi/l,4\pi/l$ and with $P=+$,  $\circ$, and $P=-$, $\bullet$. 
Solid red curves are Nambu-Goto predictions.
Black dashed line denotes lower boundary of scattering state formed of 
three (anti)fundamental flux tubes with same momentum, black solid line
of one $k=2A$ and one antifundamental flux tube.
Vertical line denotes location of `deconfinement' transition.}
\label{fig_EgsQallR84_n6f}
\end{figure}

\begin{figure}[htb]
\begin	{center}
\leavevmode
\input	{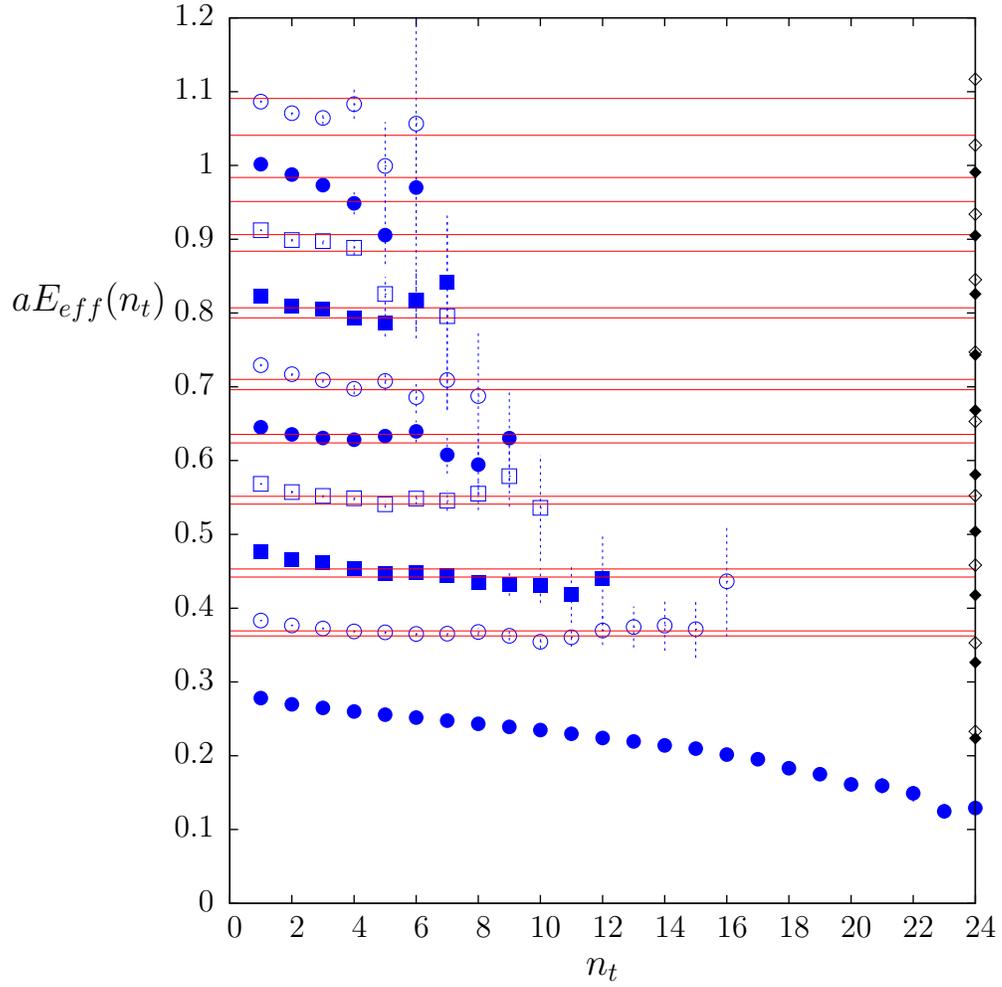}
\end	{center}
\caption{Ground state effective masses of $\underline{84}$ flux tube with 
$l/a=$16, 20, 24, 28, 32, 36, 40, 44, 48, 52.
Solid lines give error bands of our plateaux estimates. Diamonds on axis
indicate $2A+{\bar{f}}$ (solid) and $f+f+{\bar{f}}$ (open) decay thresholds.}
\label{fig_Eeffr84_n6f}
\end{figure}

%
%

\begin{figure}[htb]
\begin	{center}
\leavevmode
\input	{plot_EgsQallr120_n6f.tex}
\end	{center}
\caption{Ground states in the $\underline{120}$ representation for 
$p=0,2\pi/l$ and with $P=+$,  $\circ$, and $P=-$, $\bullet$. 
Solid red curves are Nambu-Goto predictions.
Black dashed line denotes lower boundary of scattering state formed of 
three (anti)fundamental flux tubes with same momentum, black solid line
of one $k=2A$ and one antifundamental flux tube.
Vertical line denotes location of `deconfinement' transition.}
\label{fig_EgsQallr120_n6f}
\end{figure}

\clearpage 
%
%

\begin{figure}[htb]
\begin	{center}
\leavevmode
\input	{plot_heuristic_ground_l32_K2AS.tex}
\input	{plot_heuristic_ground_l32_ADJ.tex}
\end	{center}
\caption{Overlap, as in eqn(\ref{overlap}), of $k=2A$ and adjoint 
ground states onto low-lying fundamental states. For $l=32a$.}
\label{fig_over2AAdjl32_gs}
\end{figure}

\begin{figure}[htb]
\begin	{center}
\leavevmode
\input	{plotk2aground_l64.tex}
\input	{plotadjground_l64.tex}
\end	{center}
\caption{Overlap, as in eqn(\ref{overlap}), of $k=2A$ and adjoint 
ground states onto low-lying fundamental states. For $l=64a$.}
\label{fig_over2AAdjl64_gs}
\end{figure}

\begin{figure}[htb]
\begin	{center}
\leavevmode
\input	{plot_heuristic_first_l32_K2AS.tex}
\input	{plotk2afirst_l64.tex}
\end	{center}
\caption{Overlap, as in eqn(\ref{overlap}), of $k=2A$ first excited $p=0$ 
state onto low-lying fundamental states. For $l=32a$ and $l=64a$.}
\label{fig_over2Al3264_ex1}
\end{figure}

\begin{figure}[htb]
\begin	{center}
\leavevmode
\input	{plot_heuristic_first_l32_K3AS.tex}
\input	{plotk3afirst_l64.tex}
\end	{center}
\caption{Overlap, as in eqn(\ref{overlap}), of $k=3A$ first excited $p=0$ 
state onto low-lying fundamental states. For $l=32a$ and $l=64a$.}
\label{fig_over3Al3264_ex1}
\end{figure}

\begin{figure}[htb]
\begin	{center}
\leavevmode
\input	{plot_heuristic_second_l32_K2AS.tex}
\input	{plot_heuristic_third_l32_K2AS.tex}
\end	{center}
\caption{Overlap, as in eqn(\ref{overlap}), of $k=2A$ second and third 
excited $p=0$ states onto low-lying fundamental states. For $l=32a$.}
\label{fig_over2Al32_ex2ex3}
\end{figure}

\begin{figure}[htb]
\begin	{center}
\leavevmode
\input	{plot_heuristic_second_l52_K2AS.tex}
\input	{plot_heuristic_third_l52_K2AS.tex}
\end	{center}
\caption{Overlap, as in eqn(\ref{overlap}), of $k=2A$ second and third 
excited $p=0$ states onto low-lying fundamental states. For $l=52a$.}
\label{fig_over2Al52_ex2ex3}
\end{figure}

\end{document}